\documentclass{aa}  
\usepackage{graphicx}
\usepackage{lscape}
\usepackage{rotating}
\usepackage{txfonts}
\begin{document}
   \title{The second and third parameters of the horizontal branch in globular
   clusters \thanks{Tables 1, 2, 3, 4, 5, 6, 7, 8, 10, 11, and 12, are only available
in electronic form at the CDS via anonymous
ftp to cdsarc.u-strasbg.fr (130.79.128.5) or via
http://cdsweb.u-strasbg.fr/cgi-bin/qcat?J/A+A/???/???}
}
   \subtitle{}

   \authorrunning{R.G. Gratton et al.}
   \titlerunning{The second and third parameters of the horizontal branch}

   \author{R.G. Gratton,
          \inst{1}
          E. Carretta,
          \inst{2}
          A. Bragaglia,
          \inst{2}
          S. Lucatello,
          \inst{1,3}
	  \and
          V. D'Orazi
          \inst{1}
	}
\offprints{R.G. Gratton}

   \institute{INAF-Osservatorio Astronomico di Padova, Vicolo dell'Osservatorio 
              5, I-35122 Padova, Italy\\
              \email{raffaele.gratton@oapd.inaf.it, sara.lucatello@oapd.inaf.it, valentina.dorazi@oapd.inaf.it}
              \and
          INAF-Osservatorio Astronomico di Bologna, Via Ranzani 1, I-40127
	  Bologna, Italy\\
	  \email{angela.bragaglia@oabo.inaf.it, eugenio.carretta@oabo.inaf.it}
	  \and
	  Excellence Cluster Universe, Technische Universit\"at M\"unchen, 
	  Boltzmannstr. 2, D-85748, Garching, Germany
}

\date{Received .....; accepted .....}

   
  \abstract
{The second parameter (the first being metallicity)
defining the distribution of stars on the horizontal branch (HB) of globular
clusters (GCs) has long been one of the major open issues in our understanding of
the evolution of normal stars. Large photometric and spectroscopic databases 
are now available: they include large
and homogeneous sets of colour-magnitude diagrams, cluster ages, and 
homogeneous data about chemical compositions from our FLAMES survey.}
{We use these databases to re-examine this issue.} 
{We use the photometric data to derive median and extreme (i.e., the values
including 90\% of the distribution) colours and magnitudes of stars along
the HB for about a hundred GCs. We transform these into median and extreme
masses of stars on the HB, using the models developed by the Pisa group, and taking
into account evolutionary effects. We compare these masses with those
expected at the tip of the red giant branch (RGB) to derive the total mass
lost by the stars.}
{We find that a simple linear dependence on metallicity of this total mass lost 
describes quite well the median colours of HB stars. Assuming this mass loss law to be 
universal, we find that age is the main second parameter,
determining many of the most relevant features related to HBs. In particular, 
it allows us to explain the Oosterhoff dichotomy as a consequence of the peculiar 
age-metallicity distribution of GCs in our Galaxy, although both Oosterhoff groups 
have GCs spanning a rather large range in ages. However, at least an additional - 
third - parameter is clearly required. The most likely candidate is the He abundance, 
which might be different in GC stars belonging to the different stellar generations 
whose presence was previously derived from the Na-O and Mg-Al 
anticorrelations. Variations in the median He abundance allow us to explain the 
extremely blue HB of GCs like NGC~6254 (=M~10) and NGC~1904 (=M~79); such variations 
are found to be (weakly) correlated with the values of the R-parameter (that is the 
ratio of the number of stars on the HB and on the RGB). We also show that suitable 
He abundances allow deriving ages from the HB which are consistent with those 
obtained from the Main Sequence. Small corrections to these latter ages are then proposed. We find 
that a very tight age-metallicity relation (with a scatter below 4\%) can be obtained 
for GCs kinematically related to the disk and bulge, once these corrections are 
applied. Furthermore, star-to-star variations in the He content, combined with a small
random term, explain very well the extension of the HB. There is a strong
correlation between this extension and the interquartile of the Na-O anticorrelation, 
strongly supporting the hypothesis that the third parameter for GC HBs is He.
Finally, there are strong indications that the main driver for these variations in the 
He-content within GCs is the total cluster mass. There are a few GCs exhibiting
exceptional behaviours (including NGC~104=47 Tuc and in less measure NGC~5272=M~3); 
however, they can be perhaps accommodated in a scenario for the formation of GCs that
relates their origin to cooling flows generated after very large episodes of star
formation, as proposed by Carretta et al. (2009d). }
{}

\maketitle
%

\section{Introduction}

Sandage \& Wallerstein (1960) noticed that the
distribution with colour/temperature of stars on the horizontal branch (HB)
of globular clusters (GCs) is roughly correlated with their metal content.
A few years later, this observation was explained by the first successful
models of HB stars describing the effect of metal content on the efficiency
of H-shell burning in low mass stars where He is burning in the core
(Faulkner 1966). However, soon after this important theoretical achievement,
van den Bergh (1967) and Sandage \& Wildey (1967) pointed out that the
correlation between colour/temperature and metallicity had several exceptions,
a difficulty that has become known as {\it the second parameter problem}. In the following
forty years, a large number of tentative explanations for this discrepancy
appeared in the literature, but no
overall satisfactory scenario has yet been found. A proof of the large
interest raised by this issue is that entering "globular cluster"
and "second parameter" in the ADS data base\footnote{http://adsabs.harvard.edu/}
resulted in 231 abstracts (24 since 2006) with 6031 citations on a query
made on April 15th, 2009. Of course, this search is probably
incomplete, because there are many related issues, e.g., the Oosterhoff
dichotomy in the mean periods of RR Lyrae in galactic GCs (Oosterhoff
1944; Sandage 1982), the UV upturn in the spectra of bulges and elliptical
galaxies (Code 1969), the ages and the He abundances of GCs (Iben 1968),
the mass loss law for low mass stars, which use different keywords.
The second parameter problem is certainly one of the major open issues
in our understanding of the evolution of normal stars. For
reviews of this topic, we refer to Moehler (2001) and Catelan (2009).

There are various reasons why the second parameter issue has been insofar 
so difficult to solve. The most important is that there is most likely more than 
a single second parameter. The colour of HB stars is very sensitive 
to several physical stellar quantities in the age and metallicity regime 
typical of GCs (see e.g., Rood 1973; Renzini 1977; Freeman \& Norris 1981; 
Fusi Pecci et al. 1993). Zinn (1980) and many authors after him 
convincingly showed that younger ages might explain the red colours of the 
HB of several of the outer halo GCs (see Dotter et al. 2008 for a similar
line of thought). However, progress in the 
determination of the (relative) ages of GCs (e.g., Stetson et al. 1996; Rosenberg 
et al. 1999; De Angeli et al. 2005; Mar\'in-Franch et al. 2009) demonstrated 
that this cannot be the only second parameter. The same result had previously been
obtained even more directly from the broad distribution in colours of HB stars 
within some individual GCs (see e.g., Ferraro et al. 1990). 

Since a change 
in the mass of stars on the HB may well cause even large variations in their 
colours (Rood 1973), a special mass-loss law has become a popular explanation 
(see Dotter 2008 for an example of this approach). 
Unfortunately, the physics of mass loss is very poorly known at present
(see e.g., Willson 2000; Meszeros et al. 2009; Dupree et al. 2009). Many different
mechanisms may affect mass loss (see e.g., Rich et al. 1997; Green et al.
1997; Soker \& Harpaz 2000) and empirical evidence is inadequate for fully
constraining them (see e.g., Peterson 1982; Origlia et al. 2007, 2008).
Given these limitations, it is not yet possible to build an ab-initio model
for the estimation of mass loss from GC stars. Hence we prefer to carefully restrict our
assumptions: we looked for solutions with a mass loss law based on as few 
simple parameters as possible, for which we may
obtain constraints from independent observations.

Additional second parameters considered included 
He abundances, the ratio of CNO to Fe abundances, stellar rotation or 
binarity (see e.g., Freeman \& Norris 1981), and cluster concentration 
(Fusi Pecci et al. 1993). However, all of these explanations 
were found to be unsatisfactory overall. In the most successful cases, they might 
explain some groups of stars with anomalous colours on the HB (e.g., most 
field O-B subdwarfs are binaries, see e.g., Maxted et al. 2001; 
Napiwotzki et al. 2004; Han et al. 2003; however most of these in GCs seem 
to be single stars, see e.g., Moni Bidin et al. 2006). In the least successful cases, they 
are inconsistent with the data (CNO abundances: see the case of the second 
parameter pair NGC~362 and NCC~288: Shetrone \& Keane 2000 and Catelan et 
al. 2001). The situations for rotation and He abundances are more complicated. For
rotation, after the initial enthusiasm triggered by the pioneering observations of 
Peterson (1982), the problem was found to be less straightforward. More extensive 
data sets (Deliyannis et al. 1989, Behr et al. 2000a, 2000b; Behr 2003; Recio-Blanco et al. 2002)
revealed an intricate pattern, so that it seems difficult to use direct
observations of rotation along the HB to confirm its r\^ole in the second
parameter issue. While these observations do not rule out the
possibility that rotation is indeed important, we cannot avoid concluding
that the evidence for and against remains poor (see e.g., Sweigart 2002). 
We discuss the case of He abundances in subsequent sections.

Over the years, an enormous wealth of observational data has been collected, 
not only in terms of the distribution of stars along the HB with colour, but also 
the chemical composition, the period distribution of RR Lyrae, and other 
properties of GCs. Several authors 
(see e.g., Fusi Pecci et al. 1993; Catelan et al. 2001; Recio-Blanco et al. 
2006; Carretta et al. 2007) pointed out the existence of correlations between 
global cluster parameters (such as luminosity, concentration, or Galactic orbit) 
and phenomena related to the second parameter. However, the mechanism 
connecting these global properties to the evolution of individual stars 
remained elusive until a few years ago.
 
A revised approach to the problem of the second parameter can now be developed. 
It is based on what was initially considered to be a separate characteristic
of GCs, that is the abundance anomalies observed for GCs in the light elements 
CNO, Na, Mg, and Al (see Kraft 1994 and  
Gratton et al. 2004 for reviews of this topic). Early suggestions that there 
might be a correlation between these two sets of observations were made by 
Norris (1981), and more recently by Kraft (1994) and Catelan \& de Freitas 
Pacheco (1995) and several other authors after them. However, the exact mechanism 
linking the two phenomena remained unclear. The two innovative steps taken
subsequently were:
\begin{itemize}
\item the recognition that typical GCs host at least two generations of 
stars: this is required to explain the abundance anomalies observed for main sequence (MS) 
stars by Gratton et al. (2001) and Cohen et al. (2002). These observations 
contradicted the paradigm of GCs as single stellar populations, and opened 
a new view on GC formation and evolution that we are now only beginning to
explore (see Gratton et al. 2004 for early results).
\item the understanding by D'Antona et al. (2002) that (large) variations 
in the abundance of He, which are expected to be correlated with the variations in 
Na and O and other elements, might result in large differences in the 
turn-off (TO) masses of stars of similar age: this is because He-rich stars 
evolve faster than He-poor ones and thus, at a given age, He-rich stars at TO are less massive.
\footnote{As pointed out by the referee, the knowledge that He
abundance variations lead to variations in turn-off masses is much older
(see Iben \& Rood 1970). However, D'Antona et al. were the first to relate 
variations in Na, O, and other light elements to variations in He abundances,
and then TO masses. This created the link between {\it abundance
anomalies} on the RGB and the second parameter issue. }
Therefore, similar mass losses along the red giant branch (RGB) would lead 
to HB stars of very different masses, and hence colours. Not all authors estimated
variations in the He abundances as large
as those proposed by D'Antona et al. (2002; see for instance Marcolini et
al. 2009). We note however that these small spreads in He abundance (which are
usually justified on the basis of chemical evolution models) have difficulties
in reproducing the observed spread in masses along the HB and the splitting
of the MS of NGC~2808 (Lee et al. 2005, Piotto et al. 2007).
\end{itemize}

That a combination of age and He differences may explain the second 
parameter is very attractive for 
several reasons: (i) The large variations in He and the Na-O 
anticorrelation are explained by the presence of different generations of 
stars in GCs, and it is then easy to link them to general cluster 
properties, such as their mass or location in the Galaxy, which seem to play 
a r\^ole in the second parameter issue; (ii) These different stellar 
generations may well be used to explain discontinuous and often discrete 
distributions of stars along the HB, such as that observed e.g., in NGC~2808 
(D'Antona et al.  2005); (iii) Very accurate photometric data 
detect multiple main sequences in some GCs (Bedin et al. 2004; Piotto et 
al. 2007) that can only be explained by assuming large variations in the He 
content (Norris et al. 2004; Piotto et al. 2005; Milone et al. 2010).

While these observations are extremely interesting, they are rather limited in number: 
the data discussed insofar only concern a handful of massive GCs. A more 
comprehensive study of a large set of GCs, analysed in a homogeneous 
way, was lacking until recently. Such an analysis is now 
possible, thanks to the large databases of colour magnitude diagrams (CMD) and accurate ages provided 
by ground (Rosenberg et al. 2000a,b) and space observations (e.g., Piotto et 
al. 2002), and the extensive data on the Na-O anticorrelation from our 
FLAMES survey (Carretta et al. 2009a,b and references therein), 
complemented by literature data.
In this paper, we present an exploration of these unprecedented 
databases. In the first part of the paper, we consider the evidence
provided by extensive photometric datasets, from both ground-based and
HST observations for a sample of almost a hundred GCs, deriving the properties 
of HBs as defined by their median values and extension, and examining their 
correlations with metallicity and age. This analysis produces a simple mass-loss 
law, that explains the median colours of HB stars. However, as found by 
several authors before, an additional parameter is needed to explain the HB 
colours of GCs with an extreme blue HB (BHB) and the spread of colours in many 
other cases. In the second part of the paper, we
consider variations in the He content as a possible explanation of these
discrepancies, we derive the He abundance variations required explaining the
observed properties of the HBs, and discuss the implications for MS photometry. 
In the third part of the paper, we search
for additional evidence that He is indeed the additional parameter
required to explain the HB morphology. In particular, we explore the correlations
with other chemical anomalies, namely the Na-O anticorrelation. For this purpose,
we consider a smaller but still quite large sample of 24 GCs, 
including classical second parameter cases (NGC~288 and NGC~362; 
NGC~5272 and NGC~6205), a list of blue HB clusters (such as NGC~1904 or NGC~6752), 
and the very extended HB cases (such as NGC~2808, NGC~6388 and NGC~6441). The aim 
of our discussion is to convince the reader that a combination of age and He 
abundance variations, the latter being related to multiple generations of stars within 
each GC, is a promising scenario to clarify most of the so far unexplained 
characteristics of the second parameter issue. We emphasise that we do consider
that additional effects (e.g., binarity) may affect the colour of stars 
along the HB, but the r\^oles played by the age and He abundance variations are 
probably dominant.

\section{Median and extreme colours of stars along the HB}\label{colors}

Several authors have suggested that age is the (main) second parameter determining
the colour of HB stars (e.g., Zinn 1985; Demarque et al. 1989; Lee et al. 1994). 
In this section, we present a revised evaluation of this
issue, performing an extensive comparison between observations and models. 
We exploit the large databases of globular cluster photometry available
on the web and the latest age and metallicity estimates.

\subsection{Observational data}

\begin{figure}
\includegraphics[width=9cm]{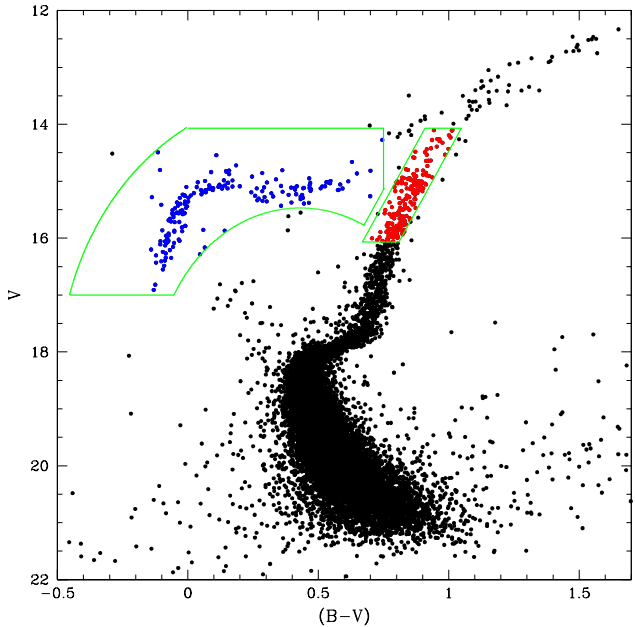}
\caption{Observed CMD for NGC~5904. RGB and HB stars are
marked with red and blue points, respectively, while the green boxes represent 
the selection region of each component.}
\label{f:select}
\end{figure}

The first step of our procedure was to derive the median and range of colours and 
magnitudes (hereinafter the range including 90\% of the stars) along the HB. 
We derived these quantities from the databases of ground-based and HST (snapshot) 
observations presented by Rosenberg et al. (2000a,b) and Piotto et al.
(2002), respectively. These databases were selected because they are publicly
available and include a large number of clusters analysed with
very homogeneous methods. The HST-ACS survey (Sarajedini et al. 2007) is 
providing new, very high quality data, which will largely supersede those older 
catalogues; unfortunately, this data set is not publicly available yet.
\footnote{An analysis of the ages of GCs and median colours of the HB based
on these ACS data has been published after this paper was written (Dotter et al.
2010). A complete comparison with our results would require a long section.
However, we note that there are very tight correlations between the V-I colours
from ACS data, and the B-V and V-I ones of this paper. The ages by Dotter et al.
also agree fairly well, at least in a statistical sense, with those considered in
this paper. Not surprisingly, there is also agreement at least in the first
major conclusion: age is the main second parameter. However, Dotter et al. suggest
cluster concentration as third parameter, while we propose absolute magnitudes,
on turn related to the He content. We note that Dotter et al. did not examine
the extension of the HB, limiting their study to the median colours. We argue
that it is not easy to discuss this third parameter using this approach. 
Furthermore, there is some correlation between luminosity and concentration
for GCs, which may justify their result. However, our analysis of the extension
of the HB shows that this is much more strongly correlated with the cluster absolute 
magnitudes, rather than their concentration.}
However, as we will see, the older photometry is of excellent quality, with 
only a few caveats. The ground database includes V and I CMDs and photometric 
tables for 52 GCs, from which we 
dropped E3, which is too scarcely populated for the present purposes, and 
$\omega$~Cen, whose spread in metal abundance makes the interpretation of data far 
more complicated than for the remaining clusters. The HST database contains 
F439W and F555W CMDs and photometric tables for 71 GCs.
In the following, we considered B and V colours obtained from the HST
photometry through the transformations given by Piotto et al. (2002). There 
are 23 clusters in common between the two samples, so data are available
for a total of 98 GCs. 

\begin{figure}
\includegraphics[width=9cm]{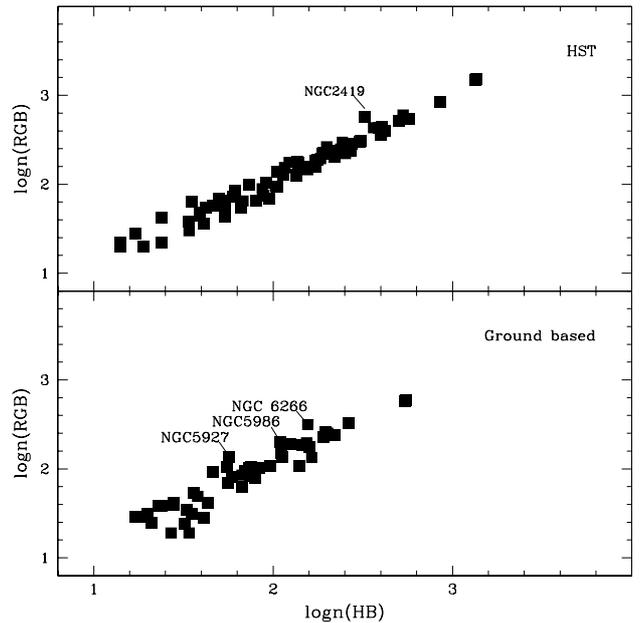}
\caption{Comparison between the number of HB and RGB stars that are 
within $\pm 1$ mag of the mean magnitude of the HB V(HB). Upper and lower panels
are for HST and ground-based data. Labels indicate the points relative to those
GCs for which star counts are not reliable, using this diagnostics.}
\label{f:nhbnrgb}
\end{figure}

Table~\ref{t:tnew0} gives basic parameters for the clusters. Metallicities [Fe/H] 
are from our re-analysis of GC metallicity (Carretta et al. 2009c; note
that we adopted a metallicity of [Fe/H]=-1.35 for NGC~6256, from Stephens \& Frogel 
2004), while apparent distance moduli $(m-M)_V$, reddenings $E(B-V)$, absolute magnitudes 
M$_{\rm V}$, and HBR are from the Harris catalogue (1996). Relative ages are from
the compilation of Carretta et al. (2009d), and are mainly from Mar\'in-Franch et
al. (2009) and De Angeli et al. (2005, this last modified to be on the same scale), 
corrected to have values consistent with the metallicities by Carretta et al. 
(2009c). Masses at both the turn-off (M$_{\rm TO}$) and the tip of the RGB (M$_{\rm RGB}$) 
were computed as explained in Sect.~\ref{sec:massloss}. IQR values are the 
interquartile of the Na-O, and Mg-Al (anti-)correlations and come from Carretta et 
al. (2009b), with a few additional data from literature (Shetrone \& Keane 2000 for 
NGC~362; Sneden et al. 2004 and Cohen \& Melendez 2005 for NGC~5272 and NGC~6205; 
Marino et al. 2008, 2009 for NGC~6121 and NGC~6656, respectively, and Yong et al. 
2005 for NGC~6752). Finally, $\log{\rm T_{eff}^{max}}$(HB) is the maximum temperature 
of HB stars, from Recio-Blanco et al. (2006), complemented by data for a few
clusters evaluated by Carretta et al.(2009d). For each cluster, we first identified 
the region covered by the HB, as well as that occupied by RGB stars that are within 
$\pm 1$ mag of the mean magnitude of the HB V(HB), taken from the on line version of 
the Harris catalogue of GCs (Harris 1996). Figure~\ref{f:select} shows an
example of the selection of HB and RGB stars. 

Completeness of the photometry can be an issue in a statistical study
such as the present one. HST-based observations are generally complete to magnitude 
$V\sim 21.5$ (with some exceptions on both the brighter and fainter side), which
is much fainter than the HB level at the RR Lyrae colour in the vast majority of
GCs of this sample. We tested the completeness of the HB in the HST photometry by 
checking whether the number of HB stars is not unexpectedly small compared to the number 
of RGB stars that are within $\pm 1$ mag of V(HB) 
(see Fig.~\ref{f:nhbnrgb}). On average, this ratio is $0.934\pm 0.020$\ 
for GCs with HST data. The only cluster for which there is a clear deficiency of
observed HB stars (a value smaller than the average one by more than twice 
the standard deviation, as estimated by Poisson statistics) is the farthest 
one (NGC~2419: Harris et al. 1996), for which this ratio is $0.56\pm 0.09$. 
The next smallest value is for NGC~6624, for which the value is $0.70\pm 0.14$, 
which is only 1.6 standard deviation below the average value. However, the 
ground-based observations are shallower, with typical limiting magnitudes in the 
range 19-20. For this reason, extreme HB stars are sometimes missing from these 
data; we then generally give preference to the HST 
snapshot data in our discussion. However, the ground-based data provide very useful 
information about a number of nearby GCs missing in the Piotto et al. (2002) 
database\footnote{The databases we considered actually also include entries for 
variable stars (on the HB mainly RR Lyrae). These entries are very inaccurate, 
since usually data at very few epochs are available, and photometry in different 
bands is not based on data acquired simultaneously. However, while inaccurate, these data still have some 
statistical meaning; for this reason, they are considered throughout our analysis.}.

\begin{figure*}
\includegraphics[width=16cm]{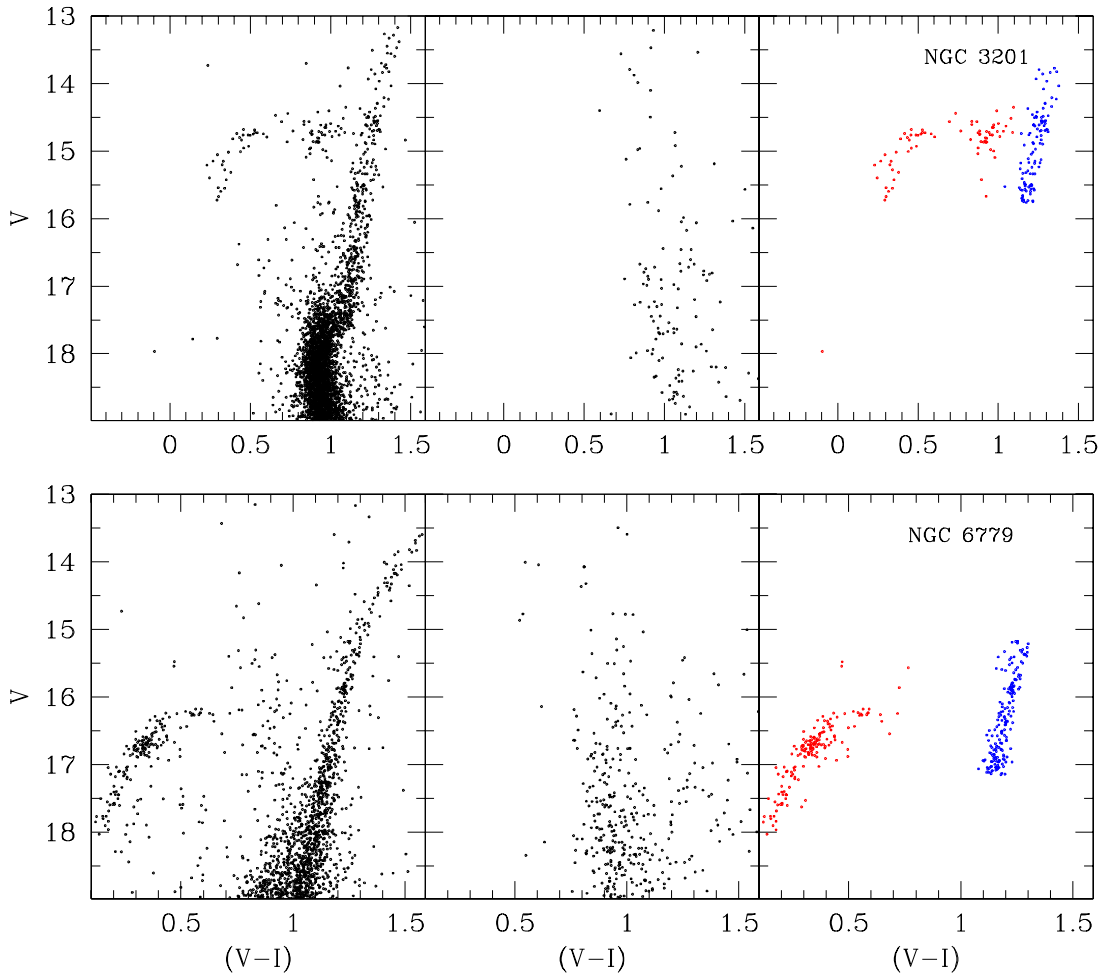}
\caption{Two examples of the application of the statistical field subtraction
used in this paper (top: NGC~3201; bottom NGC~6779); both cases are based on 
ground-based photometry. The left panels give the original photometry; the
central panels give the CMD of field stars appropriate
for the direction and reddening of the GCs, from the TRILEGAL model of the
Milky Way (Girardi et al. 2005; Vanhollebeke et al. 2009); the right panels are 
the field-subtracted HB (red points) and RGB (blue points). The RGB stars are 
only those within $\pm 1$~mag of the HB level, according to Harris (1996).}
\label{f:contamination}
\end{figure*}

\begin{figure}
\includegraphics[width=9cm]{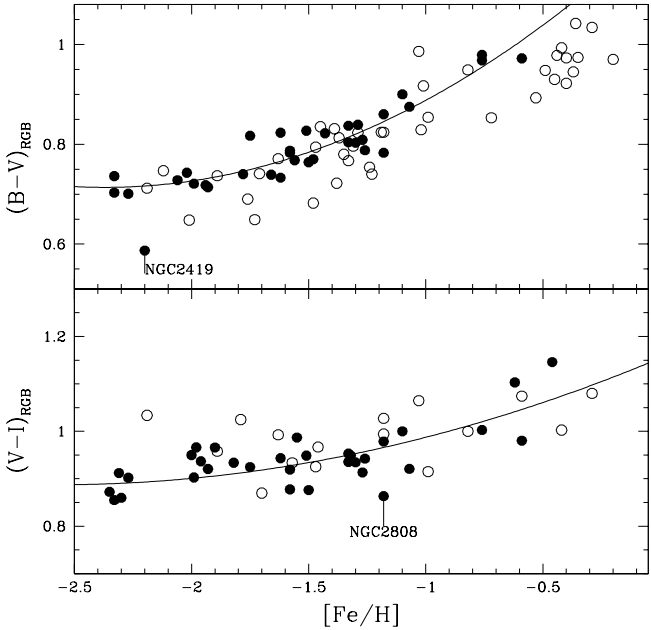}
\caption{Run of (B$-$V)$_{\rm RGB,0}$ and (V$-$I)$_{\rm RGB,0}$ with metallicity.
Open and filled symbols represent clusters with E(B$-$V) larger and smaller than 0.25,
respectively, while the solid curves show the relations given by Equations 
(\ref{eq:1}) and (\ref{eq:2}). A few points relative to interesting clusters are marked.}
\label{f:fig1}
\end{figure}

Particular care was devoted to the separation between the HB and the RGB, 
which is difficult for very metal-rich GCs when differential reddening is large. 
We also tried to minimize the impact of contamination by field stars, 
which may significantly affect the determination of the colour extremes, but
only marginally affect the determination of median colours and magnitudes, 
which are robust estimators. Unfortunately, while very extensive and homogeneous, 
the databases that we used include neither comparison fields allowing a (statistical) 
subtraction of field contaminants, nor a full membership study. Our procedure 
was to first estimate those field stars expected to fall within the region of
the CMD that we identified with the HB (Col. 3 of Table~\ref{t:tnew1}) and 
the RGB, based on the galactic model TRILEGAL (Girardi et al. 2005; 
Vanhollebeke et al. 2009). We then subtracted from the CMD the star 
closest to each field star (weighting differences in colours five times more
than differences in magnitudes); the numbers of HB and RGB stars given in this paper
are those after this subtraction. This procedure introduces some uncertainty 
in our results, in particular in cases of strongly contaminated fields. However,
eye inspection of the colour-magnitude diagrams obtained using this procedure
shows that it worked satisfactorily. Figure~\ref{f:contamination} shows a couple 
of examples of this decontamination.

We typically identified a few hundred HB stars, and a similar number of RGB 
stars. However, in a few cases we identified only a few tens of stars belonging 
to the two sequences. In these cases, larger uncertainties are associated with the 
extreme values, while the median is still quite robust (see however Sect.~\ref{spreadHB}). 

We fitted each observed HB with a polynomial (using the colour as an independent
variable), whose degree varied (between 1 and 4) from cluster-to-cluster, 
depending on the extension of the HB. For each star, we then replaced the 
observed colour and magnitude with that corresponding to the closest position 
along the polynomial. The metrics that we considered weight differences in colours five times
more than those in magnitudes. This procedure reduces the 
impact of outliers. We then ordered the stars in terms of either increasing colour or 
magnitude, and determined the median as well as the colours and magnitudes 
that include the central 90\% of the distribution.

In the case of the RGB, we determined the average colour of the sequence at
the magnitude of the HB by fitting a straight line through the points
corresponding to the RGB-selected stars; in this case, we used the magnitude
as an independent variable.

Values of observed colours and magnitudes are given in Tables~\ref{t:tnew1} 
and \ref{t:tnew2} for HST and ground-based data, respectively. 
Error bars are only those from statistics. They do not include systematic 
errors, which are due to incompleteness at faint magnitudes (most relevant for 
ground-based observations), incorrect separation of the different 
sequences (sometimes possible for most metal-rich clusters with differential 
reddening), or uncertainties in decontamination by field stars.

\begin{figure}
\includegraphics[width=9cm]{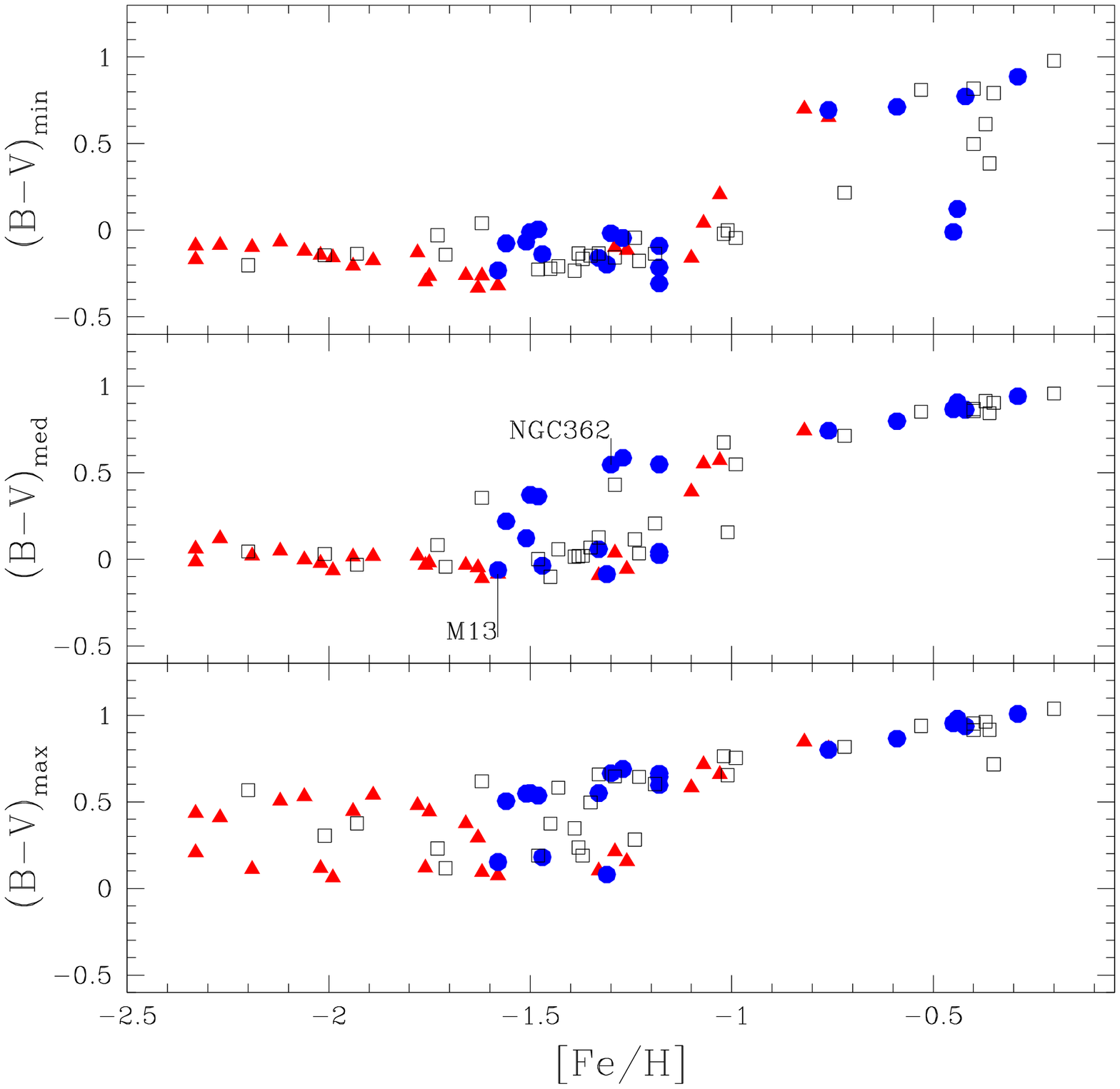}
\caption{Median and extreme colours as a function of metallicity for clusters 
observed with HST. Filled (red) triangles and (blue) dots indicate GCs older 
and younger than a relative age parameter of 0.92, respectively; clusters for 
which an age estimate is not available are marked with (black) empty squares.
Points representing a few interesting clusters are marked.}
\label{f:fig2}
\end{figure}

\setcounter{table}{0}
\begin{table*}
\caption{Basic parameters for globular clusters considered in this paper 
(M$_{\rm TO}$ and M$_{\rm RGB}$ are in solar masses). 
The complete table is available only in electronic form.}
\label{t:tnew0}      
\setlength{\tabcolsep}{1.5mm}
\begin{tabular}{lcccccccccccc}
\hline\hline
Cluster & Other & [Fe/H] & $(m-M)_V$ & $E(B-V)$ & M$_{\rm V}$ & HBR & Age & M$_{\rm TO}$ & M$_{\rm RGB}$ 
& IQR$_{\rm [Na/O]}$ &  IQR$_{\rm [Al/Mg]}$ & $\log{\rm T_{eff}^{max}}$(HB)\\
         & & dex   &    mag    &   mag   & mag           &     &    &  M$_\odot$   & M$_\odot$ 
& dex                & dex                  & K \\
\hline
NGC~104  &47 Tuc&-0.76 &13.37 & 0.04 &-9.42 & -0.99 & 0.95 & 0.862 & 0.909 &0.472 & 0.091 &3.756 \\
NGC~288  &      &-1.32 &14.83 & 0.03 &-6.74 &  0.98 & 0.90 & 0.827 & 0.869 &0.776 & 0.059 &4.221	\\
NGC~362  &      &-1.30 &14.81 & 0.03 &-8.41 & -0.87 & 0.80 & 0.858 & 0.900 &0.670 &	     &4.079 \\
IC~1257  &      &-1.73 &19.25 & 0.73 &-6.15 & -0.71 &      &       &       &      &	     &	    \\  
NGC~1261 &      &-1.27 &16.10 & 0.01 &-7.81 &  1.00 & 0.79 & 0.847 & 0.887 &      &	     &4.079 \\
\hline
\end{tabular}
\end{table*}

\begin{figure}
\includegraphics[width=9cm]{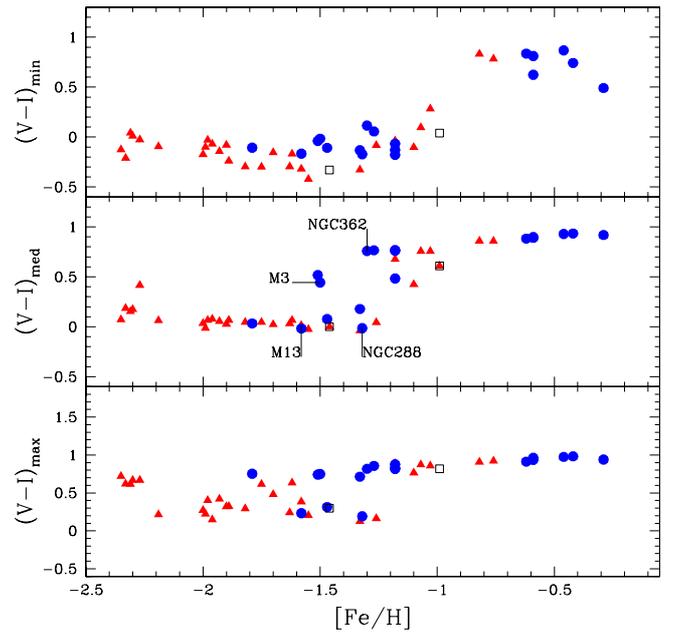}
\caption{Same as Fig.~\ref{f:fig2} but for ground-based sample.}
\label{f:fig3}
\end{figure}

\setcounter{table}{1}
\begin{small}
\begin{table*}
\caption{Photometric data from the HST-snapshot (Piotto et al. 2002).  
The complete table is available only in electronic form.}
\label{t:tnew1} 
\hskip-0.7cm
\setlength{\tabcolsep}{1.5mm}
\begin{tabular}{lccccccccc} 
\hline\hline    
Cluster & n(HB) & N$_{\rm Field}$ &
$F555_{\rm med}$ & $(F439-F555)_{\rm med}$ & $F555_{\rm min}$ & 
$(F439-F555)_{\rm min}$ & $(F439-F555)_{\rm max}$ & n(RGB) & 
$(F439-F555)_{\rm RGB}$\\ 
       &        & &
    mag          &           mag             &          mag    &
    mag                 &      mag                &         & mag\\
\hline
NGC~104 & 363&  0 & 14.062$\pm$0.002& 0.815$\pm$0.005& 14.047$\pm$0.003& 0.719$\pm$0.005& 0.896$\pm$0.081& 432& 1.067$\pm$0.004 \\
NGC~362 & 274&  0 & 15.517$\pm$0.003& 0.563$\pm$0.005& 15.900$\pm$0.168&-0.003$\pm$0.036& 0.691$\pm$0.051& 282& 0.865$\pm$0.004 \\
IC~1257 &  34&  3 & 20.440$\pm$0.063& 0.722$\pm$0.032& 20.645$\pm$0.031& 0.599$\pm$0.024& 0.889$\pm$0.028&  38& 1.522$\pm$0.013 \\
NGC~1261& 135&  0 & 16.821$\pm$0.009& 0.584$\pm$0.011& 17.324$\pm$0.129&-0.047$\pm$0.038& 0.697$\pm$0.022& 124& 0.849$\pm$0.006 \\
NGC~1851& 307&  0 & 16.205$\pm$0.017& 0.572$\pm$0.026& 16.908$\pm$0.094&-0.057$\pm$0.010& 0.705$\pm$0.088& 310& 0.919$\pm$0.004 \\
\hline
\end{tabular}
\end{table*}
\end{small}

\setcounter{table}{2}
\begin{table*}
\caption{Photometric data from ground-based data (Rosenberg et al. 2000a,b).  
The complete table is available only in electronic form.}            
\label{t:tnew2}      
\centering                          
\begin{tabular}{lccccccccc}        
\hline\hline
Cluster & n(HB) & N$_{\rm Field}$ & 
$V_{\rm med}$ & $(V-I)_{\rm med}$ &
$V_{\rm min}$ & $(V-I)_{\rm min}$ &
$(V-I)_{\rm max}$ &
n(RGB) &
$(V-I)_{\rm RGB}$\\
       &        & &
    mag          &           mag             &          mag    &
    mag                 &      mag                &         & mag\\
\hline
NGC~104 &192&  0 &  13.970$\pm$0.002& 0.892$\pm$0.006& 13.996$\pm$0.003& 0.817$\pm$0.009& 0.955$\pm$0.004& 225& 1.054$\pm$0.005 \\
NGC~288 & 83&  0 &  16.083$\pm$0.114& 0.022$\pm$0.022& 17.110$\pm$0.129&-0.136$\pm$0.016& 0.228$\pm$0.531& 101& 0.987$\pm$0.006 \\
NGC~362 & 56&  1 &  15.371$\pm$0.010& 0.779$\pm$0.015& 15.523$\pm$0.003& 0.135$\pm$0.020& 0.838$\pm$0.034&  69& 0.973$\pm$0.007 \\
NGC~1261&221&  1 &  16.653$\pm$0.003& 0.736$\pm$0.005& 16.942$\pm$0.272& 0.026$\pm$0.095& 0.826$\pm$0.021& 239& 0.926$\pm$0.004 \\
NGC~1851& 33&  3 &  16.177$\pm$0.020& 0.800$\pm$0.037& 17.086$\pm$0.143&-0.028$\pm$0.014& 0.855$\pm$0.009&  35& 1.004$\pm$0.009 \\
\hline
\end{tabular}
\end{table*}

\subsection{From observed to intrinsic colours}

Observed colours and magnitudes differ from intrinsic ones for mainly three 
reasons: (i) errors in photometric calibration and transformation to the
standard system (colour and magnitude); (ii) distance to the stars; and (iii) 
interstellar absorption and reddening. To reduce the impact of these uncertainties, 
we used the difference between observed and predicted colours of the RGB at the 
HB level ($(B-V)_{\rm RGB,0}$ and $(V-I)_{\rm RGB,0}$ for HST and ground-based 
observations, respectively) to estimate the offset in colours appropriate for each 
cluster, and then corrected the median and the extreme colours of the HB for 
these offsets. To a first approximation, errors in reddening and photometric 
calibration should cancel out when determining these quantities. However, 
$(B-V)_{\rm RGB,0}$\ and $(V-I)_{\rm RGB,0}$\ both depend on metallicity. To
estimate this dependence, we fitted with polynomials the relations obtained
for clusters with only moderate reddening ($E(B-V)<0.25$), using reddening 
from Harris (1996) and metallicities from Carretta et al. (2009c). We obtained 
the relations
\begin{equation}\label{eq:1}
(B-V)_{\rm RGB,0} = 1.235 + 0.4397~{\rm [Fe/H]} + 0.0927~{\rm [Fe/H]}^2,
\end{equation}
and
\begin{equation}\label{eq:2}
(V-I)_{\rm RGB,0} = 1.154 + 0.2066~{\rm [Fe/H]} + 0.0400~{\rm [Fe/H]}^2 
\end{equation}
for HST (33 clusters, r.m.s.=0.026 mag) and ground-based data (29 clusters, 
r.m.s.=0.039 mag), respectively (see Fig.~\ref{f:fig1}).

Inspection of the upper panel of Fig.~\ref{f:fig1} reveals that there is
a systematic trend in which $(B-V)_{\rm RGB,0}$ values for highly reddened metal-rich 
GCs are systematically below the calibration curve, when using the HST photometry. 
This discrepancy might have either of three causes (or a combination of them):
(i) metallicity or (ii) reddening for these clusters are overestimated; or (iii) 
the colour transformation at very red colours ($(B-V)>1.5$) is incorrect, producing
too red colours for these stars. Errors in reddening have no impact in our
analysis, since the procedure we adopted should cancel their effect. The impacts of
errors in the metallicity scale and colour transformations are more difficult
to trace throughout the analysis. However, in general they would lead to a small
underestimate of the masses of HB stars in these clusters; this would reduce
the sensitivity to metallicity of the mass-loss law we derive later, but should
have no other major impact on our analysis.

We estimated intrinsic colours by adding the offsets
from the mean relations (for $(B-V)_{\rm RGB,0}$\ and $(V-I)_{\rm RGB,0}$)
to the observed colours. Median and extreme of the colours are plotted 
against metallicity in Figs.~\ref{f:fig2} and \ref{f:fig3}, for HST and 
ground-based data, respectively. Different symbols are used for clusters of 
different relative ages (see Carretta et al. 2009d). In particular, we note 
that old clusters define a tight, unique (although quite complex) relation 
between metallicity and median colour of the HB. This indicates that the 
median location of the HB is uniquely determined by age and metallicity. The
relation between colours and masses of HB stars is however not linear, with 
the well-known, very strong 
sensitivity of colours on metallicities for the metallicity range $-1.6<$[Fe/H]$<-1.1$. 
This is more clearly explained in the next subsection. Younger clusters typically 
have redder median colours than older ones at a given metallicity, as expected.
Finally, clusters with no age determination roughly occupy the same region 
identified by the other clusters, but they seem to have a larger scatter. This 
is not unexpected, since data for these GCs are of poorer quality, and this is 
the reason why they lack an age estimate. We note also that much larger scatters 
are obtained for both minimum and maximum colours. We return to this point 
in the following.

\begin{figure}
\includegraphics[width=9cm]{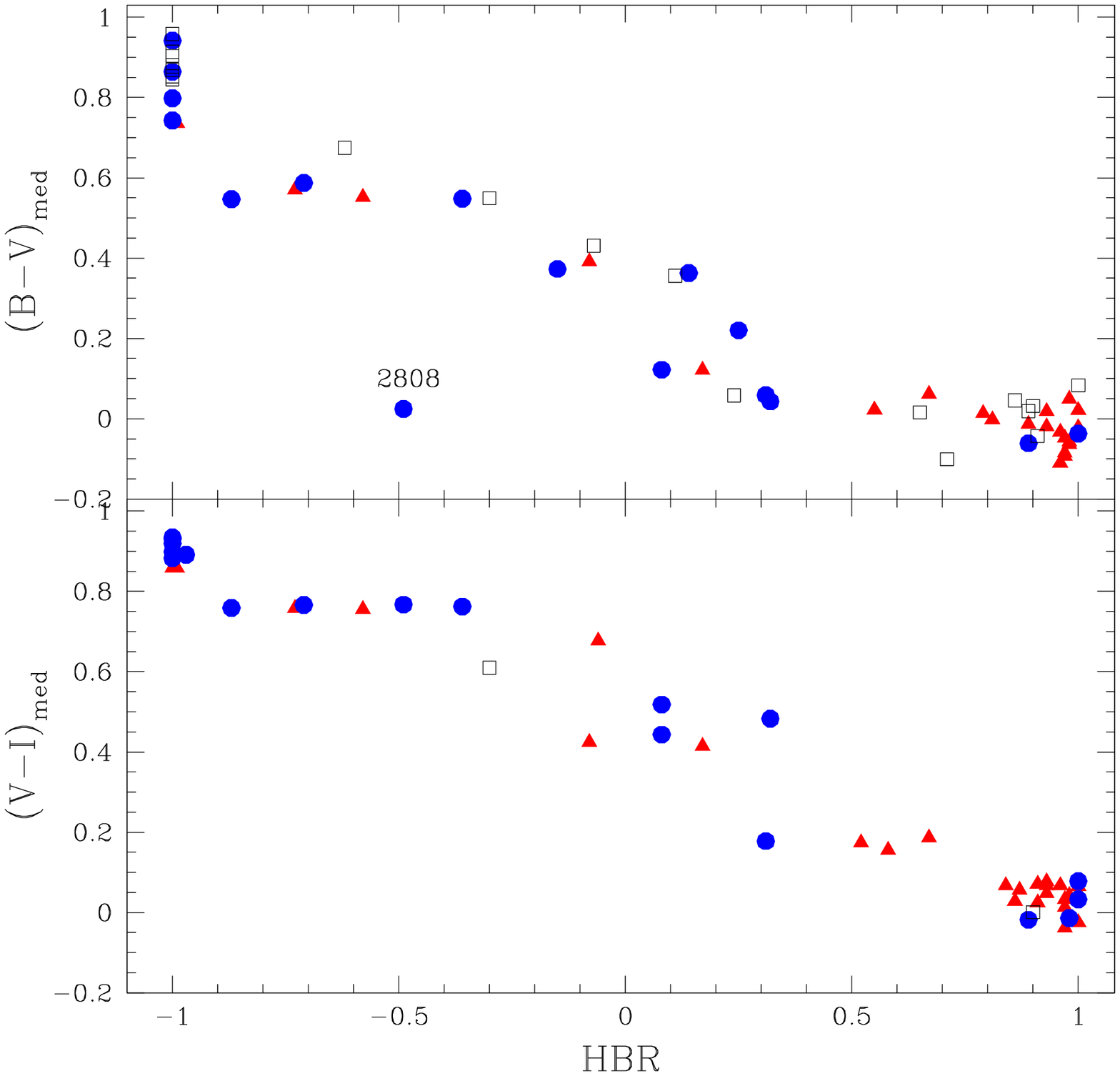}
\caption{Median colours versus HBR ratio for HST (upper panel) and ground-based 
(lower panel) clusters. Different symbols denote different ages
(see Fig.~\ref{f:fig2}).}
\label{f:fig4}
\end{figure}

\begin{figure}
\includegraphics[width=9cm]{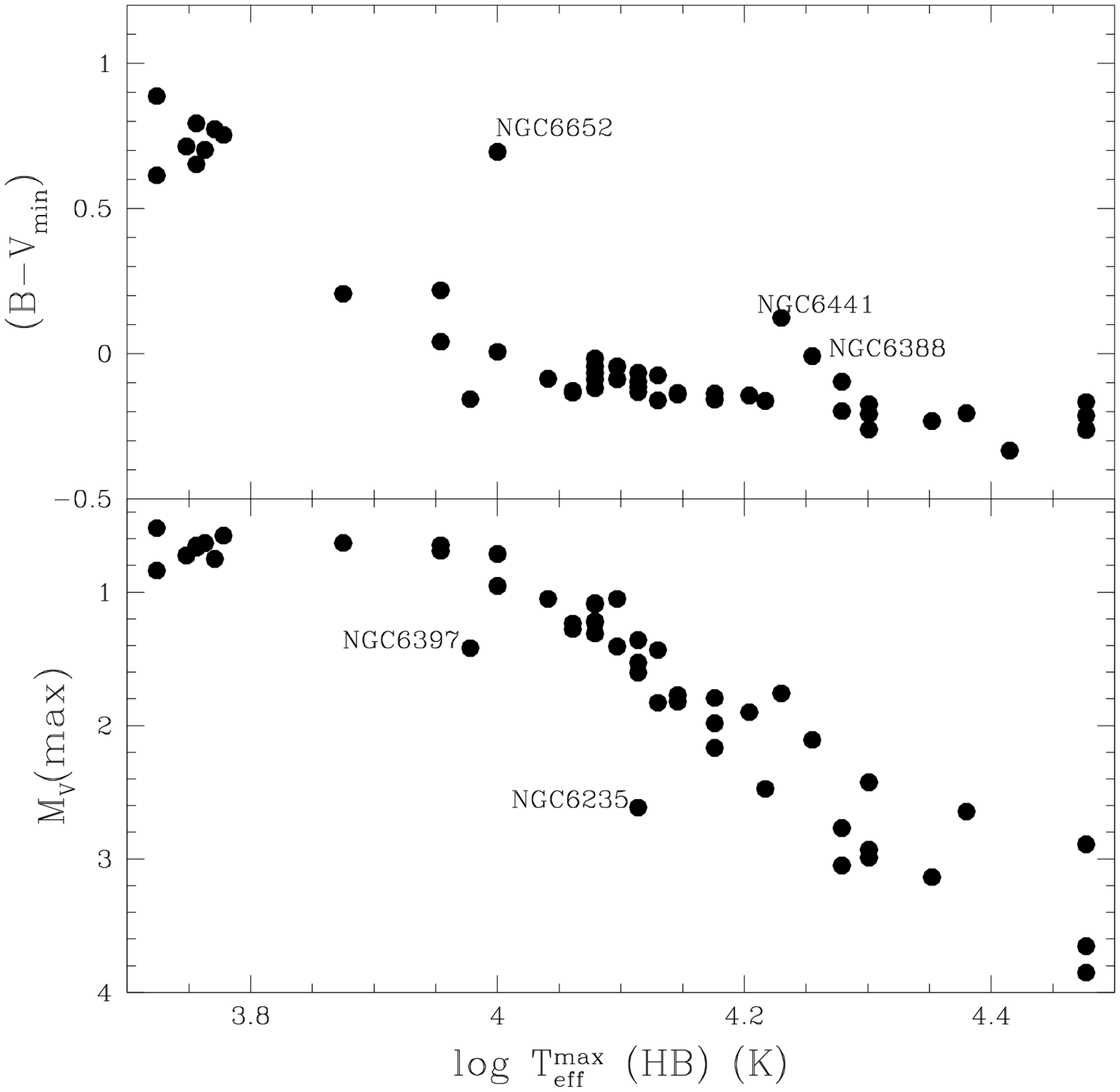}
\caption{Run of minimum colours and maximum magnitudes with the maximum temperatures
$\log {\rm T_{eff}^{max}}$(HB) retrieved by Recio-Blanco et al. (2006).}
\label{f:fig5}
\end{figure}

In the case of magnitudes, we simply corrected the observed values for the
apparent distance moduli given in Harris (1996).

We note that the plot of the median colour with metallicity is conceptually 
similar to a Lee-Zinn diagram (Lee \& Zinn 1990), since the median colours are 
very well correlated with the HBR parameter used in this diagram (see Fig. 
\ref{f:fig4}; for a definition of HBR, see Zinn 1985). The only exception to 
this close correlation is NGC~2808, when using HST (but not ground-based) data.
This is due to the multi-modal distribution of stars along the HB of this cluster, 
and to the incompleteness of the ground-based photometry, which was used to derive
the value of the HBR parameter. On the other hand, minimum colours and maximum 
absolute magnitudes along the HB are obviously well correlated with the maximum 
temperatures $\log{\rm T_{eff}^{max}}$(HB) determined by Recio-Blanco et al. 
(2006, see Fig.~\ref{f:fig5}), which were derived from the same HST photometry. 
The few discrepant cases are either GCs with poor statistics or CMDs that are 
heavily contaminated with field stars, and for which different assumptions about the 
impact of contaminants were adopted here with respect to Recio-Blanco et al. 
(2006).

Values of intrinsic colours and magnitudes are given in Tables~\ref{t:tnew3} 
and \ref{t:tnew4} for HST and ground-based data, respectively. 

\setcounter{table}{3}
\begin{table*}
\begin{center}
\begin{small}
\caption{Intrinsic HB photometric parameters for GCs with HST data.  
The complete table is available only in electronic form.}
\label{t:tnew3}  
\setlength{\tabcolsep}{1.3mm}    
\begin{tabular}{lcccccc}
\hline\hline
Cluster &($B-V)_{\rm 0,RGB}$&($B-V)_{\rm min}$ &($B-V)_{\rm med}$ &($B-V)_{\rm max}$ &$M_V{\rm max}$ 
&$M_V{\rm med}$\\ 
        & mag &     mag          &        mag        &      mag        &  mag
& mag\\
\hline
NGC~104  & 0.968 &  0.652 &  0.736 & 0.807 & 0.651 & 0.665 \\
NGC~362  & 0.803 & -0.017 &  0.547 & 0.665 & 1.090 & 0.684 \\
IC~1257  & 0.649 & -0.029 &  0.083 & 0.231 & 1.371 & 1.164 \\
NGC~1261 & 0.809 & -0.044 &  0.587 & 0.690 & 1.227 & 0.697 \\
NGC~1851 & 0.860 & -0.096 &  0.534 & 0.656 & 1.442 & 0.712 \\
\hline
\end{tabular}
\end{small}
\end{center}
\end{table*}

\setcounter{table}{4}
\begin{table*}
\begin{center}
\begin{small}
\caption{Intrinsic HB photometric parameters for GCs with ground-based data. 
The complete table is available only in electronic form.}            
\label{t:tnew4}      
\setlength{\tabcolsep}{1.3mm} 
\centering                          
\begin{tabular}{lcccccc}        
\hline\hline
Cluster &($V-I)_{\rm 0,RGB}$&($V-I)_{\rm min}$ &($V-I)_{\rm med}$ &($V-I)_{\rm max}$ &$M_V{\rm max}$ &$M_V{\rm med}$ \\
    & mag    &     mag          &        mag        &      mag        &  mag  & mag\\
\hline
NGC~104  & 1.003 &  0.783 &  0.858 & 0.921 & 0.626 & 0.600 \\
NGC~288  & 0.949 & -0.172 & -0.014 & 0.192 & 2.280 & 1.253 \\
NGC~362  & 0.935 &  0.115 &  0.759 & 0.818 & 0.713 & 0.561 \\
NGC~1261 & 0.913 &  0.056 &  0.766 & 0.856 & 0.842 & 0.553 \\
NGC~1851 & 0.978 & -0.066 &  0.762 & 0.817 & 1.616 & 0.707 \\
\hline
\end{tabular}
\end{small}
\end{center}
\end{table*}

\subsection{From intrinsic colours to masses}\label{sec:masses}

Our next step was to transform observed colours into masses of stars along the
HB. The masses were obtained by comparing the observed colours and magnitudes
with predictions from HB evolutionary models. For this purpose, we used the
database of models computed by the Pisa evolutionary group (Castellani et al. 2003, 
Cariulo et al. 2004)\footnote{http://astro.df.unipi.it/SAA/PEL/Z0.html}, 
which are particularly useful here since they provide a grid of HB
evolutionary sequences for different masses and metallicities. We first derived 
the masses appropriate for ZAHB stars of the same colours/magnitudes
of the observed loci on the HB, and then applied corrections appropriate to
taking into account the evolution of the stars away from the ZAHB.

For the first step, we derived transformations from the evolutionary 
tracks given by

\begin{eqnarray}
M({\rm ZAHB}) & = & 0.5254 - 0.0650~{\rm [Fe/H]} - 0.1181~{\rm [Fe/H]}(B-V)  \nonumber
\\
&~& +  0.1425~(B-V) - 0.6560~(B-V)^2 + \nonumber
\\
&  &  0.6277~(B-V)^3~~~~~{\rm M/M}_\odot,
\end{eqnarray}
and
\begin{eqnarray}
M({\rm ZAHB}) & = & 0.5271 - 0.0629~{\rm [Fe/H]} - 0.0851~{\rm [Fe/H]}(V-I) \nonumber
\\
&  &    + 0.1399~(V-I) - 0.6490~(V-I)^2  \nonumber
\\
&  & + 0.6134~(V-I)^3~~~~~{\rm M/M}_\odot,
\end{eqnarray}
which are valid in the range $0.54<M_{\rm ZAHB}<0.72~{\rm M/M}_\odot$\ and 
$-2.5<$[Fe/H]$<-0.6$\ (some extrapolation is required for the most metal-rich
GCs). The similar equation for the $V$ magnitude is
\begin{eqnarray}
M({\rm ZAHB}) & = & 0.5544 - 0.07286~{\rm [Fe/H]} + 0.005228~{\rm [Fe/H]}~M_V^2 \nonumber
\\
              &   & -0.04555~M_V + 0.010522~M_V^2~~~~~{\rm M/M}_\odot,
\end{eqnarray}
which is valid for $M_{\rm V}>0.8$\ and $-2.5<$[Fe/H]$<-0.6$.

The evolutionary corrections were obtained by deriving masses from median and extreme
colours of synthetic HB diagrams, obtained by taking into account evolution, with 
respect to the values appropriate for the same set of stars when on the ZAHB. We
obtained the correcting formulae
\begin{eqnarray}
M_{\rm min}({\rm ev}) & = & M_{\rm min}({\rm ZAHB})-(0.0331-0.0140~{\rm [Fe/H]} \nonumber
\\
                      &   & -0.099~M_{\rm min}({\rm ZAHB}))~~~~~{\rm M/M}_\odot,
\end{eqnarray}
\begin{eqnarray}
M_{\rm med}({\rm ev}) & = & M_{\rm med}({\rm ZAHB})-(0.0246-0.0136~{\rm [Fe/H]} \nonumber
\\     
                      &   & -0.0733~M_{\rm med}({\rm ZAHB}))~~~~~{\rm M/M}_\odot,
\end{eqnarray}
and
\begin{eqnarray}
M_{\rm max}({\rm ev}) & = & M_{\rm max}({\rm ZAHB})-[0.0356+0.1972~{\rm [Fe/H]} \nonumber
\\
& & +0.1020~{\rm [Fe/H]}^2-M_{\rm max}({\rm ZAHB})(0.18936~{\rm [Fe/H]}) \nonumber
\\
& & +0.10727~{\rm[Fe/H]}^2]~~~~~{\rm M/M}_\odot,
\end{eqnarray}
for minimum, median, and maximum mass, respectively. 

In practice, we estimated masses from: (i) colours whenever these were redder than
$(B-V)_0>0.1$\ (or $(V-I)_0>0.2$), (ii) magnitudes whenever colours were bluer
than $(B-V)_0=(V-I)_0<-0.1$, and (iii) by a weighted average of the values obtained
from magnitudes and colours when they were in an intermediate range.

\begin{figure}
\includegraphics[width=9cm]{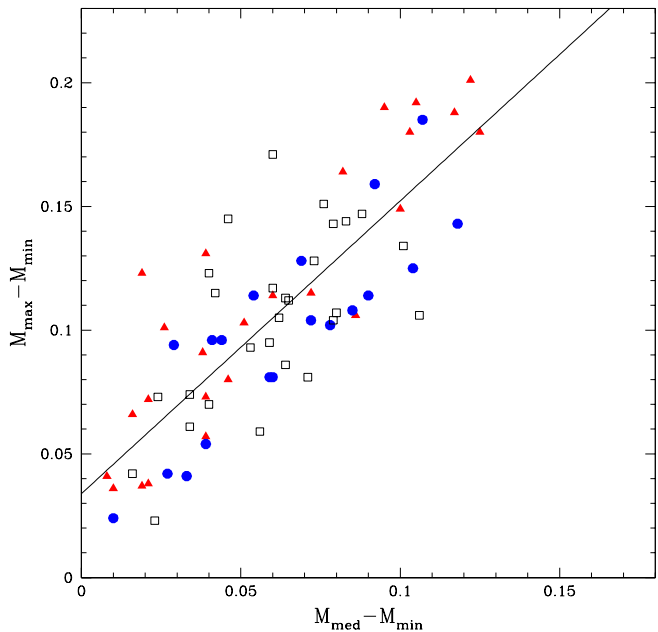}
\caption{Differences between maximum and minimum masses along the HB,
compared to differences between median and minimum masses, for clusters 
observed with HST. Filled (red) triangles and (blue) dots indicate GCs older 
and younger than a relative age parameter of 0.92, respectively; clusters for 
which an age estimate is not available are marked with (black) empty squares.}
\label{f:minmedmax}
\end{figure}

Not unexpectedly, there is a good correlation between the differences 
$M_{\rm max}-M_{\rm min}$ and $M_{\rm med}-M_{\rm min}$ (see Fig.~\ref{f:minmedmax}), 
the regression line (for HST data) being 
$M_{\rm max}-M_{\rm min}=(1.184\pm 0.097)~(M_{\rm med}-M_{\rm min})+(0.033\pm 0.026~M_\odot$
with a linear regression coefficient of r=0.82 (74 clusters). We note, however, 
that the estimate of the maximum mass is of much larger uncertainty than that
of the median mass, because of the large evolutionary effects and the impact of
smaller statistics. Furthermore, the median mass is on average much closer to the
maximum than to the minimum mass, because of the difference between these two values is 
mainly due to a nearly constant offset of about 0.03-0.04~$M_\odot$. In only a few
extreme cases (e.g., NGC~2808) is the difference between $M_{\rm med}$\ and 
$M_{\rm max}$\ sufficiently large to really affect the arguments made in
the remainder of this paper. Hereinafter we thus do not refer to $M_{\rm max}$
in our discussion.

We note that the lowest mass HB models considered here have a
mass of 0.52~M$_\odot$, the corresponding absolute magnitude predicted for these stars
being $M_V<4$. However, there are a few GCs for which the maximum absolute magnitude of
HB stars is $M_V>4$. In these cases, the masses given by our equations may be 
lower than the core masses at the He-flash for these models (Cassisi et al. 1998).
While this might be indicative of either a different origin or a late He-flash for these stars 
(see e.g., D'Cruz et al. 1996), we note that in
these cases our mass estimates are inferred from an extrapolation outside the range
covered by the stellar models, and may therefore have a larger uncertainty. The value
of the minimum mass along the HB for these GCs is certainly small, although
probably not as small as indicated by our formulae.

\setcounter{table}{5}
\begin{small}
\begin{table}
\caption{Masses on HB for GCs with HST data.  The complete table is available only in electronic form}   
\hskip-0.3cm         
\label{t:tnew5}
\setlength{\tabcolsep}{1.5mm}
\begin{tabular}{lccccr}
\hline\hline
Cluster &M$_{\rm min}$ &M$_{\rm med}$ &M$_{\rm max}$ & $\Delta$M$_{\rm max}$ &
$\Delta$M$_{\rm med}$\\ 
        & M$_\odot$    & M$_\odot$    & M$_\odot$    &  M$_\odot$            &
  M$_\odot$ \\	
\hline
NGC~104  & 0.629$\pm$0.002 & 0.648$\pm$0.001 & 0.666 & 0.286 & 0.268 \\
NGC~362  & 0.602$\pm$0.002 & 0.680$\pm$0.011 & 0.704 & 0.297 & 0.219 \\
IC~1257  & 0.615$\pm$0.011 & 0.655$\pm$0.001 & 0.685 &       &       \\
NGC~1261 & 0.593$\pm$0.001 & 0.683$\pm$0.007 & 0.707 & 0.311 & 0.222 \\
NGC~1851 & 0.579$\pm$0.003 & 0.664$\pm$0.005 & 0.687 & 0.328 & 0.243 \\
\hline
\end{tabular}
\end{table}
\end{small}

\setcounter{table}{6}
\begin{small}
\begin{table}
\caption{Masses on HB for GCs with ground-based data.  The complete table is available only in electronic form}            
\label{t:tnew6}    
\hskip -0.3cm  
\setlength{\tabcolsep}{1.5mm}
\begin{tabular}{lccccc}        
\hline\hline
Cluster &M$_{\rm min}$ &M$_{\rm med}$ &M$_{\rm max}$ & $\Delta$M$_{\rm max}$ & $\Delta$M$_{\rm med}$\\
        & M$_\odot$    & M$_\odot$    & M$_\odot$    &  M$_\odot$            &
  M$_\odot$ \\	
\hline
NGC~104  & 0.650$\pm$0.003 & 0.674$\pm$0.003 & 0.691 & 0.265 & 0.242 \\
NGC~288  & 0.570$\pm$0.001 & 0.603$\pm$0.004 & 0.645 & 0.302 & 0.269 \\
NGC~362  & 0.638$\pm$0.004 & 0.702$\pm$0.006 & 0.717 & 0.262 & 0.197 \\
NGC~1261 & 0.625$\pm$0.004 & 0.701$\pm$0.005 & 0.730 & 0.279 & 0.204 \\
NGC~1851 & 0.586$\pm$0.003 & 0.688$\pm$0.009 & 0.702 & 0.320 & 0.219 \\
\hline
\end{tabular}
\end{table}
\end{small}

Statistical errors in the mass estimates can be obtained by combining statistical
errors in the colours with the errors in the metallicities. Errors in colours 
cause an uncertainty of about 0.007~M$_\odot$ in the masses, while masses 
change by $-$0.005 M$_\odot$ if metallicity is increased by +0.1 dex in [Fe/H].

The comparison between HST and ground-based results is as follows:
\begin{itemize}
\item Minimum mass: ground-HST=$0.025\pm 0.005$M$_\odot$ (r.m.s 0.024~M$_\odot$, 27 clusters),   
where the scatter is mainly caused by GCs (NGC~2808, NGC~5986, and NGC~6266=M~62) with 
(m-M)$_{\rm V}>15.5$\ and with BHB tails that are fainter than the limiting magnitude of 
the ground-based photometry. When these three clusters are eliminated, the offset is 
$0.019\pm 0.003$~M$_\odot$ (r.m.s 0.016~M$_\odot$, 24 clusters), 
\item Median mass: ground-HST=$0.019\pm 0.003$~M$_\odot$ (r.m.s=0.019~M$_\odot$, 
28 clusters). The only clear outlier is NGC~2808. The offset is 
$0.016\pm 0.003$~M$_\odot$ (r.m.s=0.015~M$_\odot$, 27 clusters) if this cluster is 
eliminated. This confirms that the median mass is far less sensitive to outliers
and photometric errors than the minimum mass. The case of NGC~2808 is very
peculiar, because the distribution along the HB is very discontinuous
and clumpy. If the extreme blue tail were neglected (as in the case of the 
ground based photometry), the median would jump from the blue HB to the red HB,
with a change in mass of about 0.04~M$_\odot$.
\end{itemize}

\begin{figure*}
\centering
\includegraphics[width=12cm]{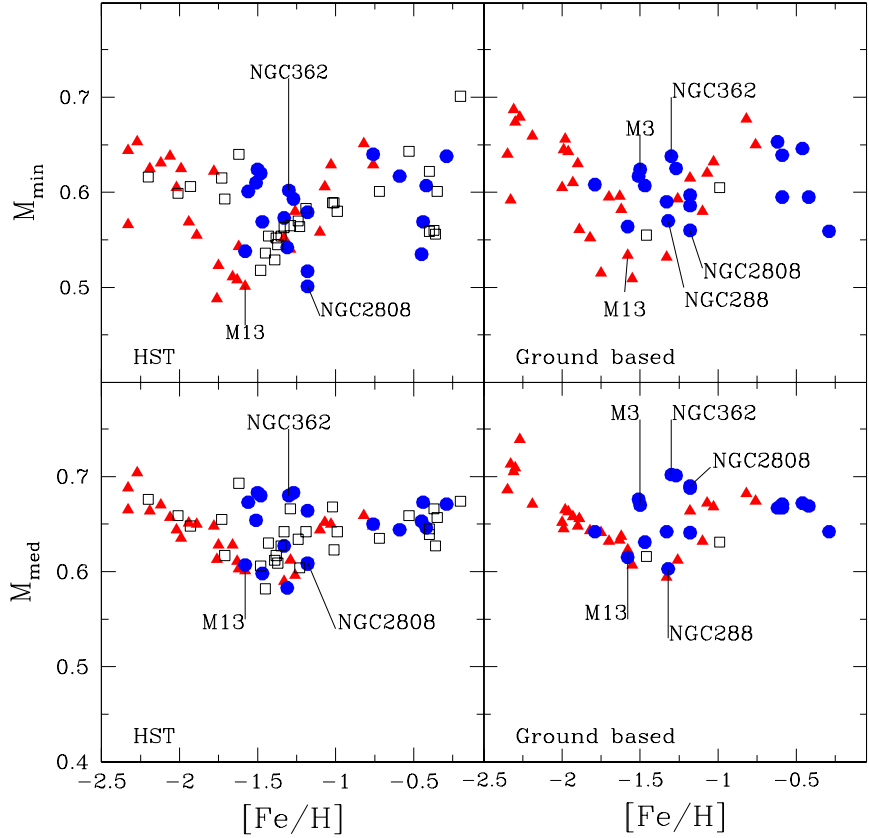}
\caption{Median and minimum masses vs. [Fe/H] for HST and ground based samples.
Symbol legend is given in Fig.~\ref{f:fig2}. A few points relative to 
interesting clusters are marked.}
\label{f:fig6}
\end{figure*}

Figure~\ref{f:fig6} illustrates the run of median and minimum masses with
metallicity [Fe/H], from HST and ground-based data, respectively. As for
colours, we indicate with different symbols clusters of different
relative ages. If we consider only the old clusters, the 
plots with median masses show a quite simple (although not linear) relation 
with metallicity, with a very small scatter about the mean line. Younger
clusters scatter above the relation defined by the old ones, as expected.
Again, clusters with no age estimates occupy a similar locus, but with a
considerable scatter.

As noted in the case of colours, when we also consider minimum masses 
the scatter is much larger than obtained for the median. We underline that 
in spite of some uncertainties in our mass derivations
this scatter is real, corresponding to a large variation in the appearance of
the HBs: we note for instance the low minimum masses that are obtained for the
GCs with very extended blue tails, such as NGC~2808 or NGC~7078. We
return to this point in Sect.~\ref{spreadHB}.

Values for the minimum, median, and maximum masses obtained following the
previous procedure are given in Tables~\ref{t:tnew5} and \ref{t:tnew6}, for HST 
and ground-based data, respectively.

\subsection{Mass-loss law}\label{sec:massloss}

The tight relation between median HB mass and metallicity for old GCs obtained in 
the previous subsection suggests that the median location of stars along the HB
of a GC can be determined using a combination of metallicity, age, and a relatively
simple and uniform mass-loss law. To show that this is indeed the case, we
first estimated the original masses of stars currently on the HB. To determine
this value, we first consider masses at TO ($M_{\rm TO}$), which can be obtained 
from their age, chemical composition, and He content, by using appropriate models. 
In practice, we used the relation
\begin{eqnarray}\label{eq:9}
M_{\rm TO} & = & 10^{-0.259 \log{\rm Age}} (0.9636+0.171~{\rm [Fe/H]} \nonumber
\\
& & +0.04073~{\rm [Fe/H]}^2)~~~~{\rm M}_\odot
\end{eqnarray}
that closely follows the isochrones of the Pisa group when Y=0.25, over the range
of parameters (age, metallicity) appropriate for GC stars. We note that Age 
in this formula is the relative age parameter of Carretta et al. (2009d), and we 
assumed that Age=1 corresponds to an isochrone age of 12.5 Gyr. We then 
corrected this value to those appropriate for stars at the tip of the RGB
(neglecting mass loss, hence assuming the mass of the stars to be identical to those at the
beginning of their evolution), using the formula
\begin{equation}\label{eq:10}
M_{\rm RGB} = M_{\rm TO} + 0.056 + 0.0117~{\rm [Fe/H]}~~~~{\rm M}_\odot,
\end{equation}
which again fits the Pisa group isochrones.

\begin{figure*}
\centering
\includegraphics[width=12cm]{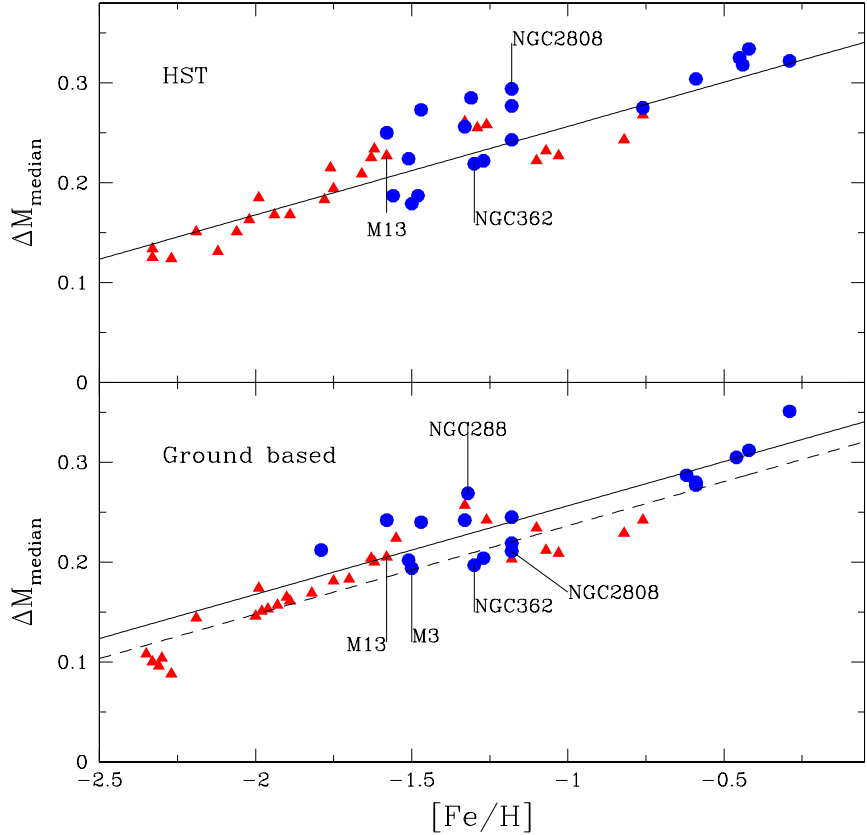}
\caption{Median mass loss as a function of metallicity for GCs older (triangles) and 
younger (dots) than 0.92. The solid line is the best-fit relation to the HST data, while the dashed one 
is just shifted by 0.02M$_\odot$, reproducing rather well the ground-based data-set.
A few points relative to interesting clusters are marked.}
\label{f:fig7}
\end{figure*}

We then assume that the original mass of stars at the tip of the RGB is equal to 
that of stars currently on the HB (on average, it should be slightly lower,
but this difference can be neglected in our discussion). The median mass lost by the
stars before reaching the HB can then be obtained by comparing their current
mass M$_{\rm med}$\ (as determined in the previous paragraph) with M$_{\rm RGB}$.
The precise evolutionary phase (before ZAHB) at which mass loss occurs is not important 
in our approach. It would of course play an important r\^ole in the comparison with mass 
loss laws. The values of the median mass loss for stars in each cluster are plotted against
metallicity in Fig.~\ref{f:fig7} for clusters with HST (upper panel) and ground-based
data (bottom panel). As usual, we used different symbols for old and young
clusters (age is required when deriving this estimate of the mass loss, so that
clusters with no age estimates cannot be plotted here). Mass losses obtained from
the two sets of data are quite similar, with a small offset of about 0.020~$M_\odot$
between them.
The best-fit straight line through the HST data is
\begin{equation}\label{eq:11}
M_{\rm lost} = (0.359\pm 0.024) + (0.0942\pm 0.0066)~{\rm [Fe/H]}~~~~{\rm M}_\odot ,
\end{equation}
which has a highly significant correlation coefficient of r=0.91 over 45 clusters. 
The scatter of points relative to each cluster around this mean line is small
(r.m.s. of 0.021 and 0.027~M$_\odot$ for old/young clusters, respectively) 
from HST data. The scatter is very similar (0.022 and 0.025~M$_\odot$ 
for old/young clusters, respectively) for ground-based data. 

At first look, this small scatter appears to be compatible with errors in 
metallicity and ages that are used in this derivation. For instance, in the case 
of the metal-rich cluster NGC~6441, changing the assumed values of the metallicity 
[Fe/H] and age parameter by 0.10 would cause variations in the values estimated 
for the mass loss by 0.014 and 0.031~M$_\odot$ (larger mass losses are obtained 
if metallicity and ages are increased). Sensitivities to these parameters are 
smaller for more metal-poor clusters; e.g., in the case of NGC~7078, similar changes 
in the assumed values of metallicity and age would cause variations in 
the mass loss of 0.007 and 0.022~M$_\odot$. In conclusion, a simple unique 
mass-loss law, combined with metallicity and age, seems able to define quite well 
the median mass and colours of HB stars for the whole sample of clusters considered 
in this analysis. However, as we will show in Sect.~\ref{ages}, the deviations of 
individual points from the mean relation are caused mostly by true physical effects.

We may compare this estimate for the total mass lost along the RGB with 
predictions by mass-loss laws in the literature. A compilation of these 
predictions is given in Table~1 and Fig.~4 of Catelan (2009). The present 
estimate of the mass lost is generally much higher than the values given by these 
predictions; only a small fraction (0.002 M$_\odot$) of this difference 
can be attributed to the identification of the mass of HB stars with that 
of stars at the tip of the RGB. The observed run compares quite well with 
a simple or modified Reimers (1975a,b) laws, or the law by VandenBerg
et al. (2000), provided that the efficiency parameter is roughly doubled. On the 
other hand, other mass-loss laws (Mullan 1978; Goldberg 1979; Judge \& Stencel
1991) provide a total mass lost along the RGB that changes too fast with
metallicity to reproduce current results.

Origlia et al. (2007) proposed a mass-loss law based on mid-IR
Spitzer observations of red giants in globular cluster. According to this
mass-loss law, GCs star should lose about 0.23~M$_\odot$ while ascending
the RGB, nearly independent of their metallicity. While this average value
roughly corresponds to what is needed to explain the HB of GCs of intermediate
metallicity, our analysis indicates that the total mass lost along the RGB
should depend on metallicity.

We note that the assumption of a particular age-metallicity relation 
(the one by Marin-Franch et al.) is implicit in our analysis. Had we adopted a different 
age-metallicity relation, the metallicity-mass-lost relation would be 
different. For instance, if the (relative) ages of disk-inner halo clusters were
independent of metallicity (rather than with a slope of $\sim -0.1$ as adopted
throughout this paper), the slope of the mass-loss law with metallicity would
be reduced by $\sim 0.02~M_\odot/dex$, from $\sim 0.09~M_\odot/dex$\ to 
$\sim 0.07~M_\odot/dex$. This would not modify significantly any of the main results 
of the paper.

\section{RR Lyrae}

RR Lyrae variables are HB stars within the classical instability strip. 
Properties of the pulsating stars may be used to constrain several basic properties, 
including mass, radius, and chemical composition. An in-depth star-by-star 
comparison, while very illuminating, requires a considerable effort, because 
individual HB stars may be within the instability strip at various phases of 
their HB evolution. This comparison then requires a synthesis of populations, 
such as those performed by Marconi et al. (2003), Di Criscienzo et al. (2004), 
Caloi \& D' Antona (2007, 2008), and D' Antona \& Caloi (2008). We defer this 
analysis to forthcoming papers; here we briefly examine only a few 
features.

\subsection{Data}
 
\begin{figure}
\includegraphics[width=9cm]{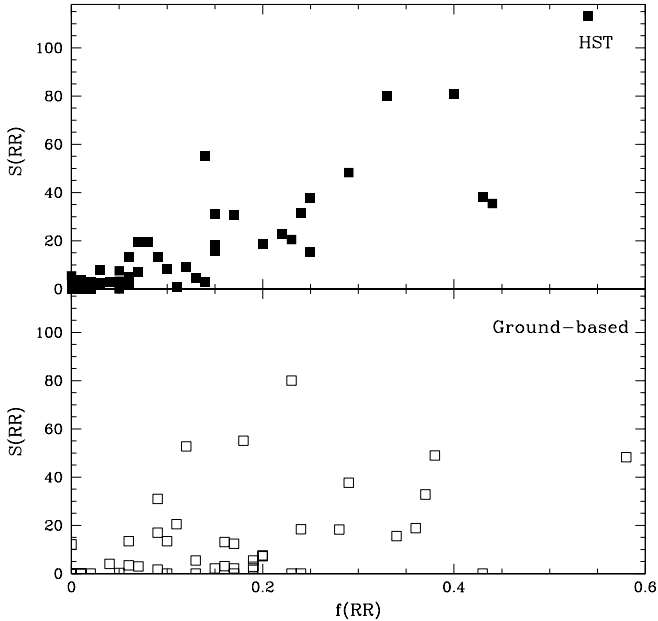}
\caption{RR Lyrae richness parameter, S(RR), as a function of the RR Lyrae 
fraction for both HST (upper panel) and ground-based sample (lower panel).}
\label{f:rr1}
\end{figure}

\setcounter{table}{7}
\begin{table*}
\begin{center}
\caption{Properties of the RR Lyrae variables for each cluster.  The complete
table is available only in electronic form.}
\label{t:rrly1}
\setlength{\tabcolsep}{1.3mm}
\centering
\begin{small}
\begin{tabular}{ccccccccc}
\hline\hline
 Cluster & S(RR)   &  n(RR)$_{\rm HST}$ & f(RR)$_{\rm HST}$ & n(RR)$_{\rm GB}$ & f(RR)$_{\rm GB}$ & f(RR) &
 $<$P$_{\rm ab}>$ &  n(P$_{\rm ab}$)\\
         &         &                    &                   &                  &                  &       &
 days              &                \\
\hline
NGC~104  &  0.2 &  2   &   0.01   &	    1	  &    0.01 &  0.01& 0.737 &    1	\\
NGC~288  &  4.0 &      &	      &	    3	  &    0.04 &  0.03& 0.678 &    1	\\
NGC~362  &  3.0 & 35   &   0.13   &	    4	  &    0.07 &  0.02& 0.542 &    7	\\
IC~1257  &      &  0   &   0.00   &	          &	        &  0.00&	   &		\\
NGC~1261 & 13.5 &  8   &   0.06   &	   21	  &    0.10 &  0.07& 0.555 &   13	\\
\hline
\end{tabular}
\end{small}
\end{center}
\end{table*}

Because of the uncertainty related to evolution, we focus only on those clusters where 
the RR Lyrae population is dominated by stars still close to the ZAHB in the
CMD. Since most of the lifetime of HB stars is spent close to their ZAHB location, 
we may be reasonably sure that this occurs for GCs that contain a large population of 
RR Lyrae. To identify these clusters, we consider the RR Lyrae richness parameter 
S(RR) as tabulated in the Harris (1996) catalogue, which is the number of RR Lyrae 
per unit cluster luminosity. \footnote{ As pointed out by the referee, the value 
of S(RR) listed in the Harris catalogue are underestimated for some GCs. A couple of 
examples are NGC~2808 (see Corwin et al. 2004) and M~62 (Contreras et al. 2005).} In 
Fig.~\ref{f:rr1}, we compared this parameter with the number of stars in the 
instability strip that we may obtain from the photometric catalogues considered in 
Sect.~\ref{colors}. We derived the number of RR Lyrae in each cluster 
by assuming that all HB stars whose dereddened $(B-V)_0$ colour is 
$0.15<(B-V)_0<0.45+0.024~({\rm [Fe/H]}+1.5)$\ are RR Lyrae (see Marconi et al. 2003 
and Di Criscienzo et al. 2004). 

In Table~\ref{t:rrly1}, we report for each cluster the S(RR) values (Col. 2) as well as 
the number and the fraction of RR Lyrae stars from both HST and ground-based samples 
(Cols. 3, 4, 5, 6). In Col. 7, we list the f(RR) estimates derived whenever possible
from S(RR) values, or otherwise from f(RR), while the mean period (P$_{\rm ab}$) and
the corresponding number of stars (n(RR$_{\rm ab}$)) are reported in Col. 8 and 9, 
respectively. These last two values are from Clement et al. (2001).

We note that the photometry considered 
throughout this paper is based on only a few, not even simultaneously acquired images in each 
filter. Hence, individual RR Lyrae will have nominal colours and magnitudes that can 
differ substantially from their mean values, and may even locate them out of our 
definitions of both the instability strip and the HB locus. However, we might expect 
that, provided that photometric errors are not too large, there should be a
correlation between the number of stars that according to the photometric catalogues 
are within the instability strip, and the true number of RR Lyrae. This appears to be
the case: the correlation between the fraction of HB stars that are nominally 
within the instability strip, and the RR Lyrae richness parameter listed by Harris, 
is quite good for HST data which have smaller photometric errors. It is somewhat 
poorer, but still fairly good for ground-based data. 
This correlation allows us to (i) calibrate the richness parameter in terms of the
fraction of HB stars that are actually RR Lyrae f(RR) (we found that this is 
represented by the mean relation $f(RR)=0.01+0.00458~S(RR)$); and (ii) complement 
the RR Lyrae richness parameter with data obtained from the photometry of those few 
clusters where this datum was missing. 

To estimate the separation between clusters whose RR Lyrae population 
is dominated by ZAHB objects rather than evolved ones, we considered in
more detail a few interesting cases. We found that using parameters that provide an accurate 
distribution of stars along the HB, the fraction of RR Lyrae stars that were 
within the instability strip also when they were on the ZAHB is 0.67 for NGC~4590 
(f(RR)=0.23), 0.45 for NGC~7078 (f(RR)=0.10), and close 
to zero for NGC~4833 (f(RR)=0.04). We note however that the values of f(RR) are quite 
uncertain, and that synthetic HBs need several parameters to reproduce various 
aspects of the colour distribution. The fractions cited above are then uncertain and 
should only be taken as indicative of a decreasing ratio of stars close to the 
ZAHB to the total of RR Lyrae variables with decreasing specific frequency; and 
that a value in the range of f(RR)$\sim 0.04-0.1$\ roughly corresponds to the 
transition between a population of RR Lyrae variables dominated by ZAHB stars to 
a population dominated by evolved objects.

Hereafter, we assume that all clusters for which $f(RR)>$0.10 (that is 
$S(RR)>$19.7) have an RR Lyrae population dominated by stars that are still
close to their ZAHB location. While as mentioned this limit is rather arbitrary, 
we found very similar results with other values in the range 0.04-0.10.

After identifying GCs for which we may assume that most RR Lyrae remain
close to the ZAHB, we can perform a sanity check of the minimum, median, and maximum
masses that we adopted for HB stars by comparing them with those derived from pulsational
properties. Table~\ref{t:dicri} lists the different mass values for the GCs analysed 
by Di Criscienzo et al. (2004). We found that masses derived from pulsational properties 
are very similar to the median
masses for the metal-rich clusters (which typically have an HB extending on both sides
of the instability strip), and closer to the maximum masses for metal-poor
clusters, which have predominantly blue HBs. While this result is unsurprising 
because we use the same HB evolutionary models as
Di Criscienzo et al. (2004), who performed a similar
comparison, at least it shows that there are no gross errors in the way we applied
these models.

\begin{figure*}
\centering
\includegraphics[width=12cm]{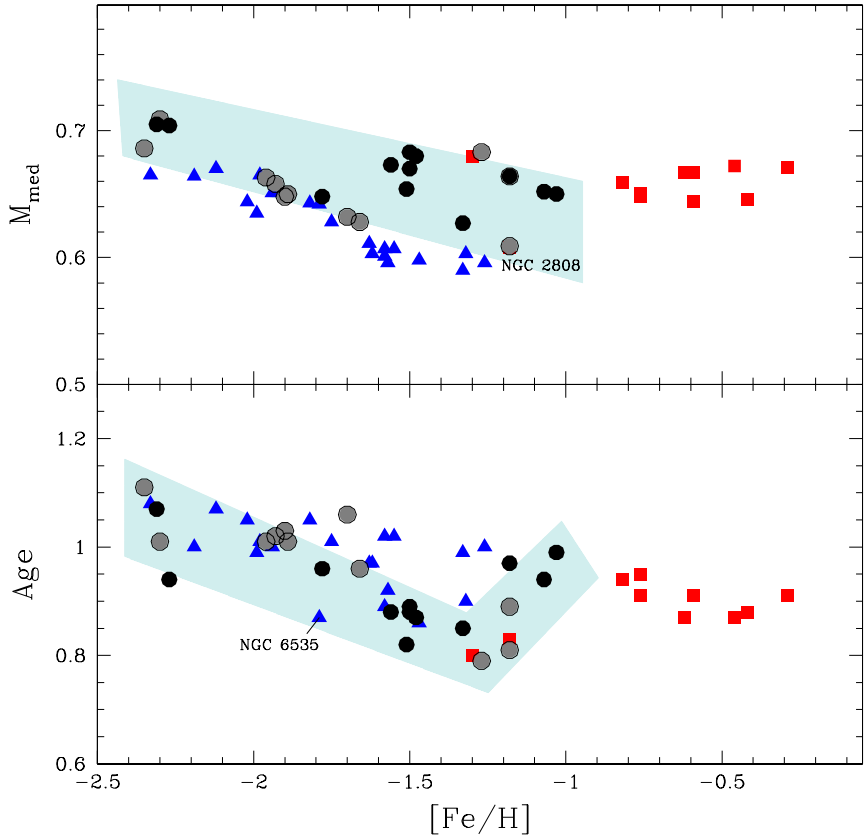}
\caption{Median mass and age as function of metallicity; different symbols in this case 
are for GCs with different RR Lyrae fractions (black: f(RR)$>$0.10, grey: 0.04$<$f(RR)$<$0.10) 
and HBR ratios (blue triangles: HBR$>$0, red squares: HBR$<$0). The shaded area is the strip 
containing GCs rich in RR Lyrae.}
\label{f:rr2}
\end{figure*}

\setcounter{table}{8}
\begin{table}
\caption{Comparison of median and maximum masses with the masses for RR Lyrae variables 
derived by Di Criscienzo et al. (2004)}
\label{t:dicri}
\centering
\begin{tabular}{lccr}
\hline\hline
NGC~& M$_{\rm RR}$ & M$_{\rm med}$ & M$_{\rm max}$\\
     & M$_\odot$    & M$_\odot$     &  M$_\odot$ \\
\hline
1851 & 0.66 & 0.666 & 0.688 \\
3201 & 0.69 & 0.654 & 0.706 \\
4499 & 0.70 & 0.693 & 0.733 \\
4590 & 0.80 & 0.704 & 0.756 \\
5272 & 0.69 & 0.651 & 0.698 \\
5466 & 0.74 & 0.685 & 0.738 \\
5904 & 0.66 & 0.627 & 0.687 \\
6362 & 0.66 & 0.652 & 0.686 \\
6809 & 0.69 & 0.638 & 0.676 \\
6934 & 0.70 & 0.673 & 0.705 \\
7078 & 0.77 & 0.688 & 0.767 \\
7089 & 0.66 & 0.628 & 0.699 \\
\hline
\end{tabular}
\end{table}

\subsection{Discussion}

One of the most intriguing characteristics of HBs is the dichotomy in the
distribution of the GCs in terms of the mean periods of variables pulsating in the
fundamental mode (RRab), discovered by Oosterhoff (1944). We note that
while the Oosterhoff dichotomy applies not only to GCs, but also to field stars in
our Galaxy (e.g., Szczygiel et al. 2009), populations of RR Lyrae sampled in other 
(smaller) galaxies such as the Magellanic Clouds or the dSphs have mean periods that are
intermediate between the two Oosterhoff groups (Oo{\sc i} and Oo{\sc ii}, e.g., 
Pritzl et al. 2004, Catelan 2009). A quite extensive discussion of various 
properties of GCs in relation to the Oosterhoff dichotomy can be found in Sect. 6 
of Catelan (2009). The information collectively is quite difficult to interpret, because 
of significant uncertainties in many of the basic parameters considered (in particular, 
ages, but also metallicities), of small number statistics, and because average periods
within a cluster depend on a complex combination of factors, mainly related to 
RR Lyrae possibly belonging to different evolutionary phases. In spite of this,
many authors have searched for a simple explanation of the Oosterhoff dichotomy, and
several of them suggested that this might be caused by a combination of ages and
metallicities, the OoII group being attributable to old, metal-poor GCs, and the
OoI to a population of young, metal-intermediate clusters (see e.g., Lee \& Carney 1999). 
We used our data to test whether this interpretation of the Oosterhoff dichotomy
is consistent with the HB scenario we consider in this paper. We
first replotted (upper panel of Fig.~\ref{f:rr2}) the metallicity-median mass 
diagram of Fig.~\ref{f:fig6}, but this time using different symbols for GCs 
that are either rich or poor in RR Lyrae: black and grey circles are for 
f$\geq$0.10 and 0.04$\leq$f$<$0.10, respectively. The GCs with f$<$0.04 are 
divided according to the HBR ratios, (blue) triangles and (red) squares 
representing HBR greater and smaller than 0, respectively. Not unexpectedly, RR Lyrae-rich 
GCs occupy a well defined locus in this diagram, roughly represented by a diagonal 
strip; GCs below this strip have blue HBs, while those above the strip have 
red HBs. We note that the anomalous behaviour of NGC~6535, being a BHB cluster with 
very low RR Lyrae fraction and located within the instability strip
(shaded box) probably reflects a small amount of data in the CMD of this cluster.  

As shown by Fig.~\ref{f:fig6}, this diagram is essentially an age-metallicity 
diagram (with some caveats that are described more clearly in the next section). 
Given the distribution of GCs in age and metallicity, there are very few galactic 
GCs with young ages and -2$<$[Fe/H]$<$-1.6. This lack of GCs with suitable parameters 
divides the galactic population of RR Lyrae rich GCs into two ensembles, which can be 
identified with the Oosterhoff groups, confirming what had already been proposed by many other 
authors. However, ages of both groups have a much larger spread than given by this 
simple consideration. This is because of two results: (i) the diagonal strip defined by 
the RR Lyrae-rich clusters has a considerable width, much larger than that caused simply by 
the range in colour of the instability strip itself; this indicates that there is 
an intrinsic spread in the masses of ZAHB stars, which is a basic property of HBs that 
we exploit throughout this paper. This is important in particular for the case of
metal-poor clusters, which are able to produce an RR Lyrae rich population over a
rather wide range of ages. (ii) Given the interplay between the original mass and the
mass-loss variation with metallicity, the strip occupied by RR Lyrae-rich clusters 
overlaps twice with the locus occupied by old GCs in the metallicity-median mass 
diagram. This implies that there are old RR Lyrae-rich GCs that are either metal-poor 
(the classical Oo~{\sc ii} clusters), or in a restricted range of metallicity around 
[Fe/H]=-1. Hence, the most metal-rich among the Oo~{\sc i} GCs belong to the old 
population, and are actually as old as the most metal-poor Oo~{\sc ii} clusters. 
This is clearly shown in the age-metallicity diagram shown in the bottom panel of 
Fig.~\ref{f:rr2}, where again we plotted with different symbols clusters that are 
either rich or poor in RR Lyrae.

This description only refers to those GCs that are very rich in RR Lyrae. There 
are many GCs with only a few RR Lyrae: as mentioned above, in these cases (generally 
metal-poor clusters with predominantly BHB) most variables can be identified with 
stars that have evolved off the ZAHB. In these cases, we 
should expect long mean periods, and the cluster to be identified as OoII (e.g., Lee 
et al. 1990). Furthermore, there are peculiar cases (including e.g., NGC~6388 and 
NGC~6441) where the RR Lyrae are actually HB stars with masses much lower than 
expected for their metallicity and age (see next section). All these cases complicate 
the simple picture described above, as discussed by Catelan (2009).

To summarise, although the Oosterhoff dichotomy indeed depends on the peculiar
distribution of galactic GCs in the age-metallicity plane, and most of the
Oo~{\sc ii} clusters are old while most of the Oo~{\sc i} clusters are young, it
is not possible to identify tout-court the Oosterhoff groups with groups of GCs having different
ages (a similar conclusion was drawn by Catelan 2009).  
							      											      
\section{Ages from median HB colours}\label{ages}							      
											      
After demonstrating that the median colour of the HB is determined mainly by		      
cluster metallicity and age, provided a suitable mass-loss law is adopted, we		      
attempt to derive ages for GCs from their median HB colour. 	      
This procedure is almost identical to that considered by several authors		      
(e.g., Lee et al. 1994), but it is now applied to a large database of GCs in a uniform way.    

\begin{figure}
\includegraphics[width=9cm]{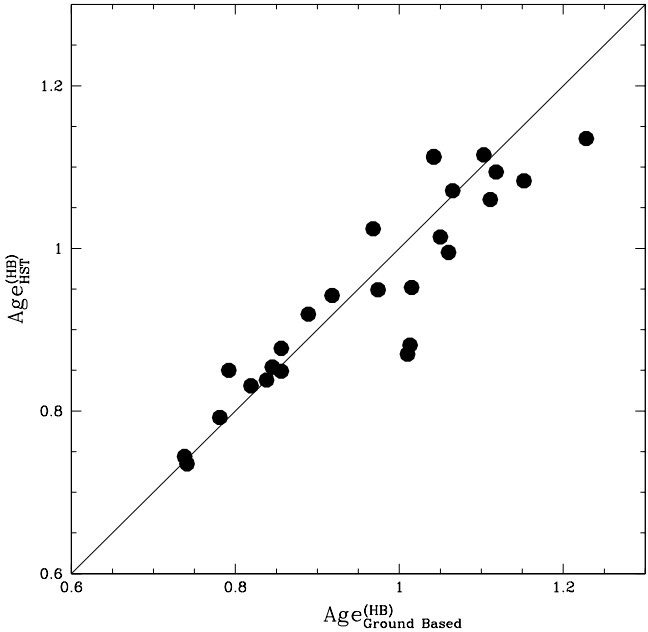}
\caption{Comparison of ages derived from HB for HST and ground-based samples.}
\label{f:fig8b}
\end{figure}

\begin{figure}
\includegraphics[width=9cm]{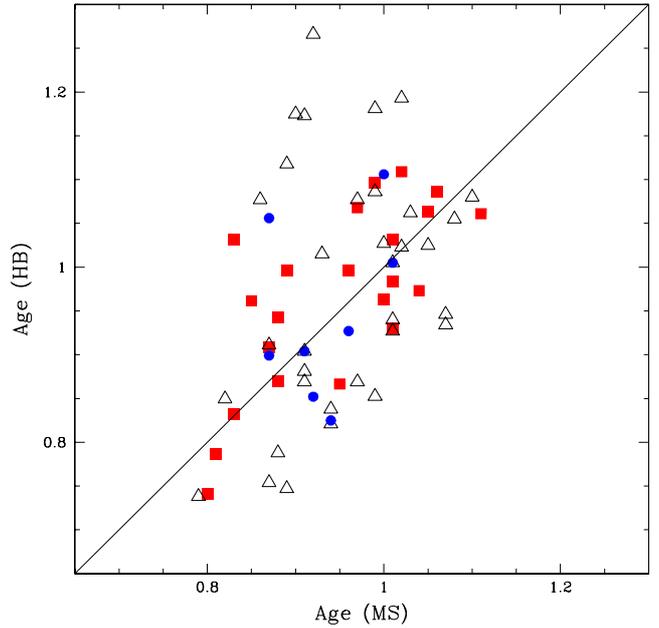}
\caption{Comparison of ages from HB and from main sequence for both HST and 
ground based samples; filled squares and circles are GCs with M$_{\rm V}$ $<$ $-$8 mag
and  M$_{\rm V}$ $\geq-$6.5, respectively, while empty triangles 
are for $-8.0 \leq$ M$_{\rm V}<$-6.5. Note that the scatter 
around a 1:1 correlation is larger for clusters of intermediate luminosity (see 
text).}
\label{f:fig8}
\end{figure}

These ages can be obtained by simply adding to the median mass of the HB stars		      
the mass lost during the previous evolution (Eq.~\ref{eq:11}), then subtracting 
the mass difference between stars at the tip of the RGB and the TO (Eq.~\ref{eq:10}), 
and finally inverting the age-metallicity-TO mass relation (Eq.~\ref{eq:9}). 
Of course, since the mean mass-loss law adopted throughout this paper
(a linear dependence on [Fe/H]) was obtained by comparing median masses
on the HB with those at the tip of the RGB, where mass loss is neglected, on
average the two sets of ages should agree. However, any difference in the ages
derived from the HB and RGB for a given GC indicates peculiarities for that GC. Taking
an extreme view, were not age the main parameter affecting the HB
morphology, we might have no correlation (or even an anticorrelation)
between MS and HB ages, only their mean values being in agreement. The derivation
of ages from HB is then crucial to understanding the second parameter issue.
These ages (expressed in our usual relative scale) are listed in Cols. 3 and 4 
of Table~\ref{t:ages} for HST and ground-based data, respectively. We compared these 
two age estimates in Fig.~\ref{f:fig8b}. The two sets of ages agree very well 
each other, with the exception of NGC~2808, for which the ground-based HB age 
is underestimated because the limiting magnitude of the relevant photometry is 
too bright to include the faint end of the HB. Once this cluster is dropped, the 
mean difference is 0.003$\pm$0.009 (r.m.s=0.046); assuming the same error in both 
age estimates, this implies an internal error of 0.032 for each of them. In the 
following, we adopt as our most robust estimates of the HB ages those determined from 
HST data, whenever available, otherwise we use values obtained from ground-based data.

\setcounter{table}{9}
\begin{table}
\caption{Clusters' ages from main sequence fitting and from HB both for HST
and ground based data-set. Y$_{\rm med}$\ is given in Table~\ref{t:tnew7}.  
The complete table is available only in electronic form.}
\label{t:ages}
\setlength{\tabcolsep}{1.5mm}
\begin{tabular}{lcccc}
\hline\hline
Cluster & Age(MS) & Age(HB) & Age(HB) & Age(Y) \\
        &         & HST     & ground  &      \\ 
\hline
NGC~104  &  0.95  & 0.877 & 0.856 & 0.922  \\
NGC~288  &  0.90  &   	 & 1.175 & 0.958  \\
NGC~362  &  0.80  & 0.744 & 0.738 & 0.789  \\
IC~1257  &        & 0.880 &	     & 	      \\
NGC~1261 &  0.79  & 0.735 & 0.741 & 0.780  \\
\hline
\end{tabular}
\end{table}

In Fig. \ref{f:fig8}, we compare these HB ages with those determined from the 
main sequence. The agreement is fairly good: the r.m.s. of the difference in 
ages from HB and from the main sequence is 0.094 for HST and 0.105 for ground-based 
data respectively: this is approximately equivalent to 1 Gyr. While this 
agreement may appear satisfactory, it is actually much larger than internal 
error bars.\footnote{This indicates that the scatter about the mean mass-loss 
relation in Fig.~\ref{f:fig7} is indeed real, as mentioned in 
Sect.~\ref{sec:massloss}, and not simply due to observational errors.} 

\begin{figure}
\includegraphics[width=9cm]{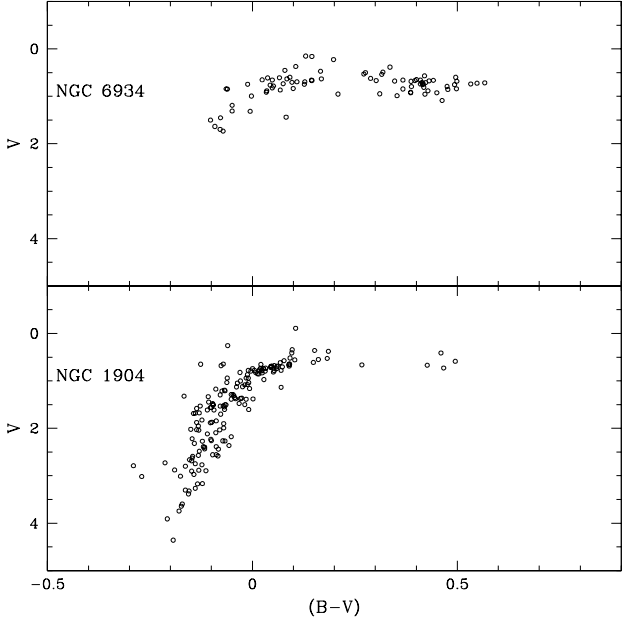}
\caption{Dereddened HST CMDs for stars on the HB of NGC~6934 and NGC~1904.}
\label{f:1904}
\end{figure}

\begin{figure}
\includegraphics[width=9cm]{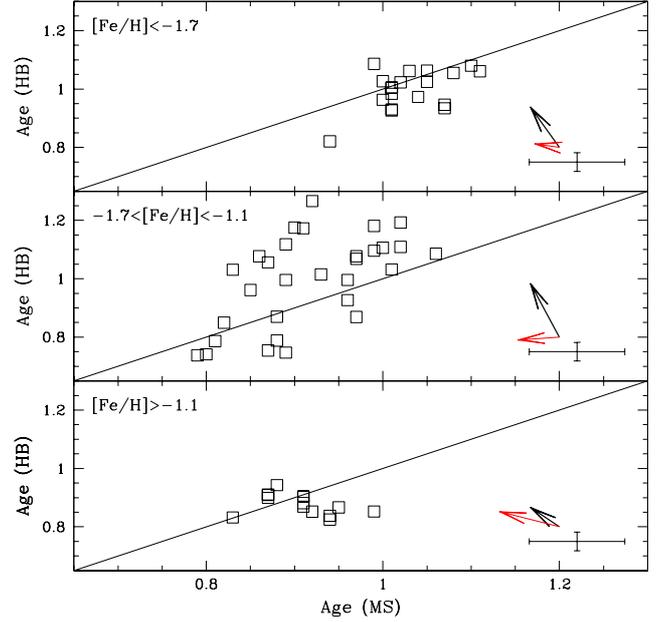}
\caption{Comparison of ages from HB and from main sequence for the whole sample 
(HST+ground based); typical error bars are shown at bottom right corner 
of each panel. The thick (black) arrows show the impact of a 0.02 variation
in the assumed He content, while the thin (red) ones represent the consequence 
of a 0.1 dex error in [Fe/H].}
\label{f:fig8c}
\end{figure}

\begin{figure}
\includegraphics[width=9cm]{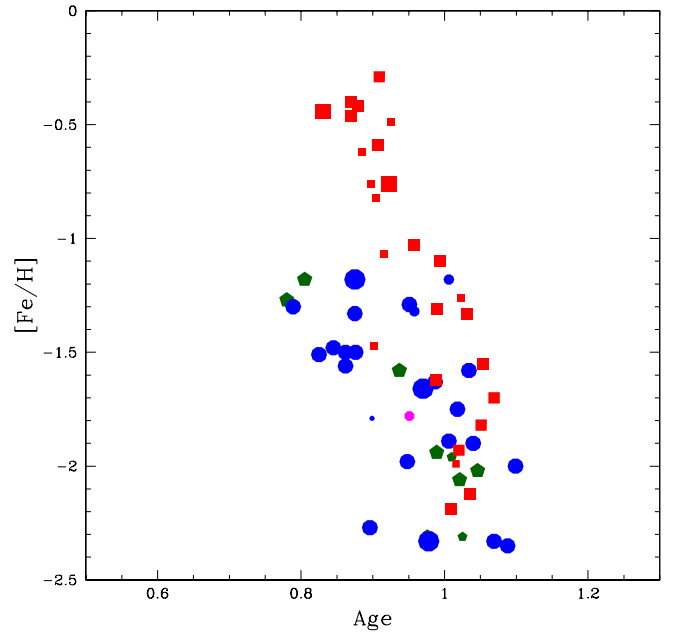}
\caption{Age-metallicity relation for different groups of globular clusters:
outer halo clusters (green pentagons), inner halo clusters (blue circles),
disk/bulge clusters (red squares). Magenta circles are GCs associated with dSphs.
Different symbol sizes are used for clusters of different luminosity.}
\label{f:fig8d}
\end{figure}

There are various ways to convince ourselves that this is indeed the case. First, we 
compared in Fig.~\ref{f:1904} the HBs of two GCs (NGC~1904 and NGC~6934) 
that have very similar metallicities ([Fe/H]=$-$1.58 and [Fe/H]=$-$1.56, 
respectively) and relative ages from MS (0.89 and 0.88). It is clear that their 
HBs are vastly different. The median mass values we obtain are 0.606$\pm$0.003 
and 0.671$\pm$0.008 for NGC~1904 and NGC~6934, respectively. The two clusters 
also have a very different frequency of RR Lyrae, i.e., f(RR)=0.02 and 0.38.  

Second, as shown by Fig.~\ref{f:fig8}, the scatter of ages from MS and from HB 
is much larger for GCs of intermediate luminosity and smaller for both brighter 
and fainter clusters. Furthermore, there is a clear trend of this scatter with 
metallicity. This is shown in Fig.~\ref{f:fig8c}, where GCs of different 
metallicity are plotted separately in the three panels. In this figure, we also 
plotted the typical error bars as well as arrows showing the impact of errors in 
the assumed metallicity and He content. These errors take consistently
into consideration the impact of the He content on the MS colour, the TO point,
and the HB. While the uncertainties in the assumed 
metallicity have a relatively moderate impact on these age estimates, those in 
the He content very significantly affect the age estimates from HB. This effect 
is particularly large at intermediate metallicity, which is where we found the 
largest discrepancies between the two sets of ages. This suggests that the He 
content might be responsible for these discrepancies (although we cannot 
exclude that other terms, e.g., the total CNO abundances, might contribute). 
We also note that the errors in the ages from MS and HB due to He are 
anticorrelated with each other: if the true He abundances are higher than we assumed, 
the MS ages are underestimated, while the HB ages are overestimated. This result 
could be exploited to simultaneously derive the values of Y$_{\rm med}$ and age 
for each cluster, by ensuring agreement between these two age derivations. The age 
values, Age(Y), obtained by applying this procedure are listed in the last 
column of Table~\ref{t:ages}. Age(Y) values are typically close to MS ages, the 
largest differences being for a few clusters with very blue HBs, such as NGC~6254
and NGC~1904. 

In principle, errors in Age(Y) should be obtained by quadratically summing 
the internal errors in MS and HB ages; the mean quadratic value determined in 
this way is 0.063. To confirm the reliability of this uncertainty estimate, 
we plot the age-metallicity relation for different groups of GCs in Fig~\ref{f:fig8d}: 
outer halo GCs, inner halo GCs, disk/bulge GCs, and GCs associated with dSphs 
(for the classification of the GCs in these different populations see Carretta et 
al. 2009d). Briefly, we replot the same age-metallicity diagram, as previously 
done in Carretta et al. (2009d) but with the new age values, i.e. Age(Y): we found 
that for the disk/bulge GCs the relation becomes tighter, with an r.m.s around a 
least squares fit of 0.036 with respect to 0.042 previously derived (note that
assuming the age values directly retrieved from Mar\'in-Franch et al. 2009,
the scatter around the fit line is $\sim$0.050). This improvement also supports 
the identification of this third parameter with variations in the median He 
abundances. Furthermore, the very small dispersion, well within the above cited 
error of 0.063, suggests that the observational uncertainties in the ages from 
MS's were most likely overestimated. We consider these Age(Y) values to be the most robust estimates 
of the relative ages currently available: they incorporate updated photometric 
data and metallicities, and take into account cluster-to-cluster variations in 
the He content. We however postpone to a future paper a full discussion of the 
age-metallicity relation for the different Galactic populations of GCs.

\section{He from HB and R-parameter}\label{sec:rparameter}

\setcounter{table}{10}
\begin{table}
\caption{R' parameter as derived from both HST and ground based data (note the
different definition with respect to the usual R parameter). The final adopted  
value $<$R'$>$ is the weighted average between the two estimates, when available.
Y($<$R'$>$) is the corresponding He abundance.  
The complete table is available only in electronic form. }
\label{t:r}
\hskip-0.3cm
\setlength{\tabcolsep}{1.5mm}
\begin{tabular}{lcccc}
\hline\hline
 NGC~/IC     & R'(HST)      & R'(ground based)& $<$R'$>$     & Y($<$R'$>$)     \\
\hline
  104 & 0.69$\pm$0.07 & 0.68$\pm$0.09 & 0.69$\pm$0.07 & 0.253$\pm$0.012 \\
  288 &               & 0.69$\pm$0.14 & 0.69$\pm$0.14 & 0.250$\pm$0.024 \\
  362 & 0.75$\pm$0.08 & 0.63$\pm$0.15 & 0.75$\pm$0.08 & 0.260$\pm$0.012 \\
 1257 & 0.49$\pm$0.14 &               & 0.49$\pm$0.14 & 0.213$\pm$0.024 \\
 1261 & 0.62$\pm$0.10 & 0.69$\pm$0.09 & 0.62$\pm$0.10 & 0.238$\pm$0.017 \\
\hline 										      
\end{tabular}										      
\end{table}										      
				
Are there other arguments supporting the idea that the third parameter required 
to explain the median colours and masses of HB stars is indeed the He content?
Is there any available evidence on variations in the median He content between 
different GCs? The most classical derivation of He abundances for GCs is obtained 
using the so-called R-parameter method (Iben 1968). The R-parameter is the
ratio of the number of HB stars, N$_{\rm HB}$, to the number of RGB stars brighter
than the HB in $M_{\rm Bol}$, N$_{\rm RGB}$. This ratio depends on the He content, 
because the higher the He content, the brighter the HB, the smaller the number of 
RGB stars brighter than this level, and hence the larger the value of R. The R-parameter can be
calibrated using models that include the variation in the lifetime of RGB 
and HB stars with metallicity. The most updated calibration of R is that by
Cassisi et al. (2003). This calibration was used by Salaris et al. (2004)
to produce R values and average He abundances for 57 GCs based on the same 
HST photometric database considered here. Salaris et al. (2004) set an upper
limit to the r.m.s. scatter of Y among these GCs of 0.019. By itself, this
upper limit to the r.m.s. scatter is fully consistent with the r.m.s. scatter 
of 0.014 for $Y_{\rm med}$\ that we derived by matching HB and MS ages. It would 
be rewarding to identify some degree of correlation between the Y values of individual
GCs derived by the two techniques. However, the correlation is actually poor,
probably because the Y values, derived using the R-method, have large error bars.
This is mainly due to limited statistics, though in a few cases other error
sources (contamination by field stars, differential reddening, photometric limits)
contribute to the noise in Y determinations.

We note that the statistical error in the R parameter can in principle be
reduced by adopting a different definition. With the usual definition, 
on average R$\sim 1.5$, there are more stars on the HB than on the RGB
brighter than the HB, which is then a major contributor to the random noise.
However, we may use a different definition of R, where the error caused by
the small number of stars on the RGB is substantially reduced, but
has a similar dependence on the helium content than the R-parameter. In practice,
we may define $R'=N_{\rm HB}/N'_{\rm RGB}$, where $N'_{\rm RGB}$ is the number of
stars on the RGB brighter than V(HB)+1. Roughly, $N'_{\rm RGB}$ is twice the value of $N_{\rm RGB}$, 
so that $R'=N_{\rm HB}/N'_{\rm RGB}\sim 0.7$. It is then clear
that the statistical errors in R' are smaller than in R. Furthermore, checks of the
stellar models of Bertelli et al. (2008) indicate that R' has roughly the same 
(fractional) dependence on Y as R, with only one important modification: the RGB
bump is brighter than V(HB)+1 in all GCs considered in this paper. Hence,
R' has a simpler (roughly linear) dependence on [Fe/H], unlike R. We
then decided to repeat the test using Y values inferred from R', rather than R.

Table~\ref{t:r} gives the values of R' deduced from HST and ground-based data;
since completeness is more of a concern for this second set of photometry, we give preference
to the HST values, whenever possible. When making this derivation, we simply used 
the HB level listed by Harris (1996). The errors related to individual entries of 
Table~\ref{t:r} are simply those from statistics, and do not include the effects of 
uncertainties in decontamination by field stars, differential reddening, and 
photometric limits. In a few cases, these effects are known and can be taken into 
consideration. On this basis, we exclude from our estimates the ground-based 
observations for NGC~2808, NGC~5986, and NGC~6266, and the HST data for NGC~2419, 
where the photometry is not deep enough to reach the extreme hot/faint tail of the 
HB. For NGC~5927, the ground-based photometry is also not good enough to discriminate 
between HB and RGB stars. The R' values are underestimated in these cases, and we did not 
consider them in our discussion.

\begin{figure}
\includegraphics[width=9cm]{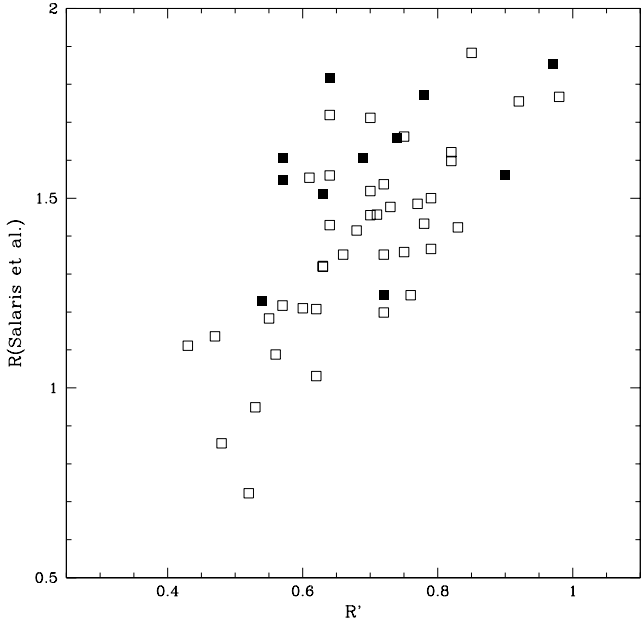}
\caption{The R' values plotted against the R values from Salaris et al. (2004); filled 
and empty squares are for GCs with [Fe/H] larger and smaller than $-$1.0 dex, respectively.
Note the different definitions of R' and R (see text),}
\label{f:rsalaris}
\end{figure}

In Fig.~\ref{f:rsalaris}, we compare the average values of R' with the values of
R obtained by Salaris et al. (2004). There is a good correlation between these
two sets of data, which improves if we consider the offset between metal-rich
and metal-poor clusters caused by the luminosity of the RGB bump; however, we
recall that this new determination is not completely independent of the Salaris
et al. one, since both use the HST snapshot photometry. A reasonable good 
correlation is also obtained with the older determinations of R by Sandquist et 
al. (2000) and Zoccali et al. (2000). 

\begin{figure}
\includegraphics[width=9cm]{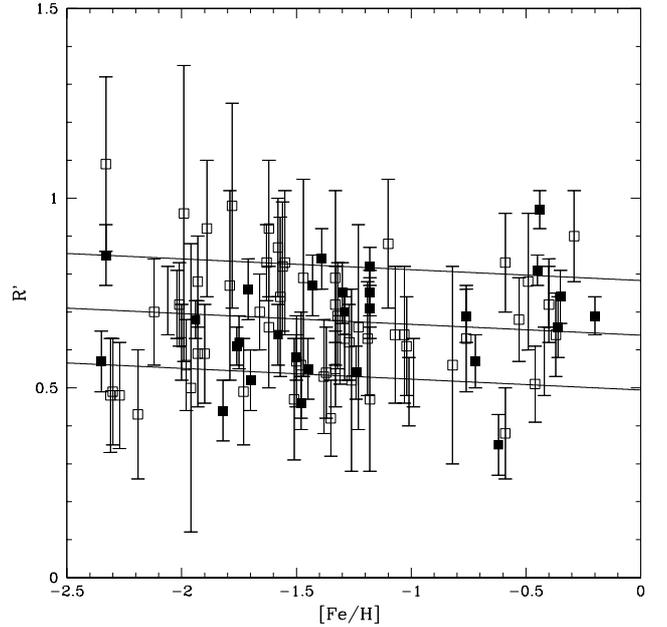}
\caption{The R' values plotted against [Fe/H]; filled and open symbols are
GCs with or without age estimate, respectively. The solid lines correspond to Y=0.225, 0.250, 0.275.}
\label{f:rfeh}
\end{figure}

The last column of Table~\ref{t:r} gives the value of Y that we could derive from these
R'. These were obtained by scaling the calibration of Cassisi et al. (2003) for
the ratio of $N'_{\rm RGB}$ to $N_{\rm RGB}$, and taking into account that 
the RGB bump is brighter than V(HB)+1 in all GCs, while being fainter than
V(HB) for GCs with [Fe/H]$>-1.1$. In Fig.~\ref{f:rfeh} we have plotted the
values of R' against the metallicity [Fe/H], and compared them with the calibrations
we used. We note the absence of an obvious bump at [Fe/H]$\sim-1.1$, which is present
in the case of the R-parameter.

\begin{figure}
\includegraphics[width=9cm]{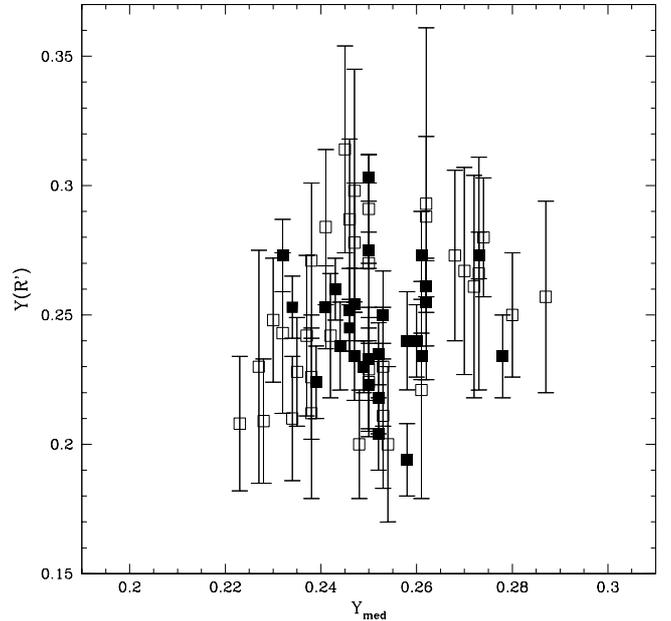}
\caption{The values of Y$_{\rm med}$ are plotted against Y(R'); filled and 
open squares are for errors smaller and larger than 0.02, respectively.}
\label{f:ymedyr}
\end{figure}

In Fig.~\ref{f:ymedyr}, we plotted the values of $Y(R')$ obtained following 
this procedure against the median He abundance $Y_{\rm med}$ required by placing 
ages estimated from the HB in agreement with those determined from the MS. 
There is some correlation between the two determinations of Y (linear 
correlation coefficient r=0.23 over 64 GCs, which is significant at a
level of confidence of $>95$\%). The scatter is clearly not 
negligible, as expected based on the rather large errors in Y(R'). However,
we note that the meaning of these two quantities is not necessarily the same.
To show this, we note that the strongest contribution to a $\chi^2$ test 
performed by comparing the two sets of He determinations is from NGC~6388 and NGC~6441;
in both these clusters, Y(R') is far larger than Y$_{\rm med}$ than
the (statistical) errors. The large value of Y(R') is explained by 
HB at the RR Lyrae colour being very bright in these clusters (and indeed
the RR Lyrae have anomalously long periods: Pritzl et al. 2002, 2003). On the
other hand, the RR Lyrae are most likely in the tail of the distribution of high
He abundances in these clusters, whose HB is predominantly red. In these cases,
Y(R') is then expected to be larger than $Y_{\rm med}$, as is indeed observed.
We also note that a much stronger correlation exists between 
Y(R') and the average of Y$_{\rm med}$ and Y$_{\rm max}$. In this case, the
correlation coefficient is r=0.294 for the same 64 GCs, which is significant
at a level of confidence of $\sim 99$\%.

We conclude that the hypothesis that the He abundances obtained by matching
ages from MS and HB is compatible with current estimates of the He
abundances from the R-method, but that more extensive data sets and more
careful comparisons (taking into consideration other possible
indication of the He abundance, see e.g., Catelan et al. 2009) are needed 
to definitively settle this issue.

\section{The colour and mass spread along the HB}\label{spreadHB}

\subsection{Observational data}

The previous discussion suggests that at least three parameters (metallicity, 
age, and {\it most likely} helium content) are required to provide a detailed explanation of the 
median colours of the HBs. We intend to demonstrate that this selection 
of parameters successfully explains the extension of the HBs. 

\begin{figure}
\includegraphics[width=9cm]{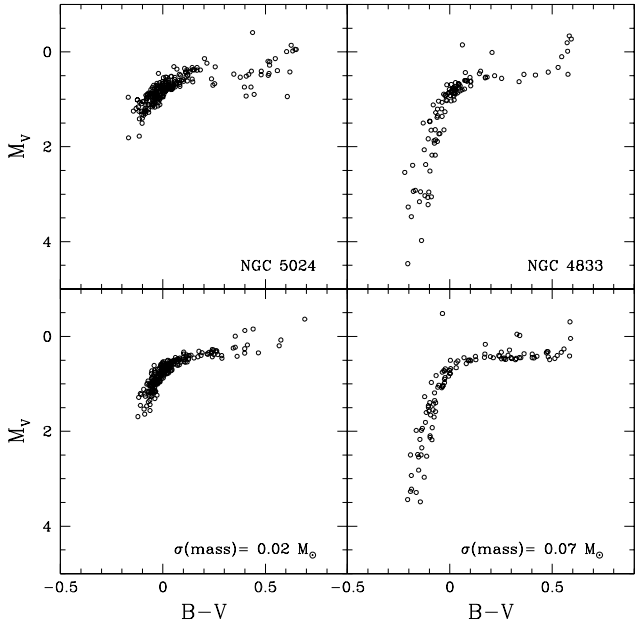}
\caption{Observed CMDs for NGC~4833 and NGC~5024 (upper panels) and synthetic diagrams 
lower panels), obtained with a Gaussian-like distribution mass loss with standard 
deviation of 0.02 M$_\odot$ and 0.07 M$_\odot$, respectively.}
\label{f:fig9}
\end{figure}

\begin{figure}
\includegraphics[width=9cm]{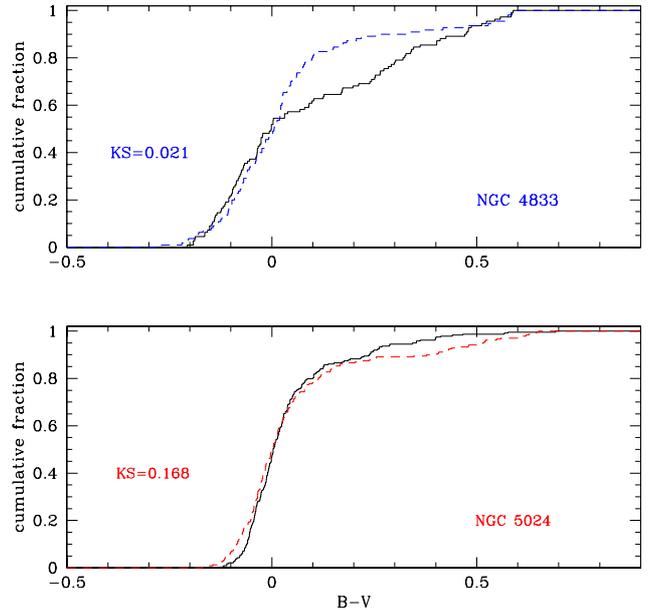}
\caption{Cumulative distribution of (B$-$V) colours from 
colour-magnitude (dashed) and synthetic (solid) diagrams for NGC~5024 and 
NGC~4833; the resulting probabilities from Kolmogorov-Smirnov (KS) test are also shown.}
\label{f:fig10}
\end{figure}

To illustrate our approach, in Fig.~\ref{f:fig9} we compare the $BV$ colour 
magnitude diagrams of NGC~4833 and NGC~5024. These two clusters have very similar ages 
(1.01 and 1.04, respectively) and metallicity ([Fe/H]=-1.89 and -2.06 respectively), 
similar values of both median colour and mass of stars on the HB (0.655 versus 0.658 
M$_\odot$), but yet very different HBs. The main difference between these two 
clusters is in the minimum mass of stars along the HB (0.561 vs. 0.649 M$_\odot$). 
We can explain the short HB of NGC~5024 in terms of a small star-to-star scatter in mass 
loss (see e.g., Rood et al. 1973, and many others since). The bottom left panel of 
Fig.~\ref{f:fig9} shows a synthetic HB obtained by assuming that the mass lost during 
the RGB phase is distributed like a Gaussian, with a standard deviation of 0.02~M$_\odot$ (this 
is the minimum value required to reproduce the shortest HBs). While this synthetic 
HB matches well the one observed in NGC~5024, it is clearly a poor match to that of 
NGC~4833, because it fails to describe its very extended faint tail. The difference between 
the two clusters is not simply in terms of the spread of mass loss. This is shown in the 
bottom right panel of Fig.~\ref{f:fig9}, where we plotted a synthetic HB obtained assuming a 
mass loss spread of 0.07 M$_\odot$, required for an HB extending as faint as the
observed one for NGC~4833. However, this synthetic HB has by far too many red stars, as
shown by a comparison of the cumulative distributions in colours (see Fig.~\ref{f:fig10})
\footnote{Small excesses of red stars along the HB - much smaller than found for 
NGC~4833 - are found often in our comparisons with synthetic HBs. This same problem 
was found by other similar analysis (see the discussion in Castellani et al. 2005). 
It might indicate either a deficiency of the adopted HB models, or a non-Gaussian 
distribution of the {\it random} terms of the mass loss, perhaps because it would be 
more appropriate to consider a Gaussian distribution in the mass-loss efficiency parameter 
(D'Cruz et al. 1996).}. A Kolmogorov-Smirnov test shows that the observed colour 
distribution disagrees with that of the synthetic HB (at about 98\% level of confidence). 
In contrast, a much better agreement is obtained between the observed and synthetic 
HBs for NGC~5024. We may conclude that at least in the case of the pair NGC~4833-NGC~5024, 
a third parameter other than age and metallicity is required, and that this parameter 
mainly affects the minimum mass along the HB.

\begin{figure}
\includegraphics[width=9cm]{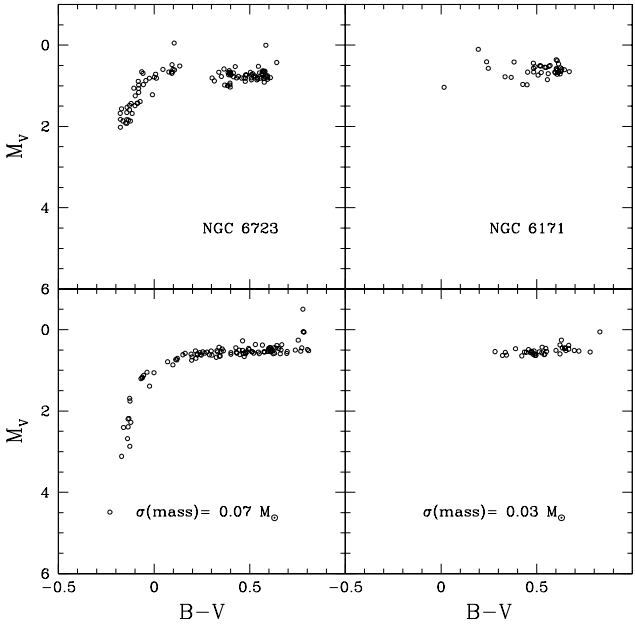}
\caption{Same as Fig.~\ref{f:fig9}, but for the GCs NGC~6171 and NGC~6723. The 
synthetic diagrams (lower panels) were obtained with a Gaussian-like distribution 
mass loss with standard deviation of 0.03 M$_\odot$ and 0.07 M$_\odot$, respectively.}
\label{f:fig11}
\end{figure}

\begin{figure}
\includegraphics[width=9cm]{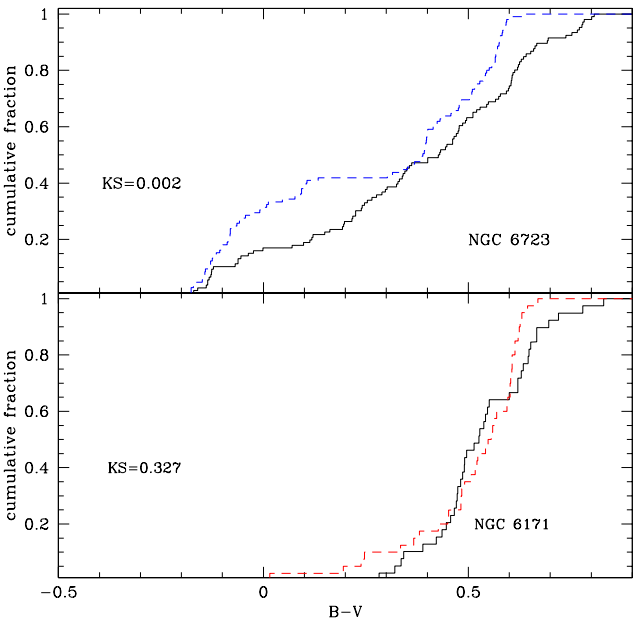}
\caption{As for Fig.~\ref{f:fig10}, cumulative distributions and KS values
for NGC~6171 and NGC~6723.}
\label{f:fig12}
\end{figure}

A similar comparison can be made using more metal-rich clusters. For this purpose,
we selected the pair NGC~6171-NGC~6723, which also have very similar metallicities
and ages ([Fe/H]=-1.03 and -1.10, Age=0.99 and 1.01, respectively), similar
median mass on the HB (0.650 and 0.644~M$_\odot$), but again vastly different
HB, due to a very different value for the minimum mass (0.629 vs. 0.558~M$_\odot$).
We can reproduce the observed HB of NGC~6171 by a
synthetic HB with a small Gaussian spread of 0.03~M$_\odot$ in the mass lost 
(see Fig.~\ref{f:fig11}).The shape of the colour 
distribution of NGC~6171 is not perfectly reproduced, again because the synthetic 
diagram predicts too many red stars, as shown by Fig.~\ref{f:fig12}: however, given 
the small size of the sample, the result of the Kolmogorov-Smirnov test returns a 32.7\% 
probability that observed and synthetic HBs are drawn from the same population, which 
is not significant. On the other hand, the distribution of colours of HB stars along 
the HB of NGC~6723 is clearly bimodal (see also Rood \& Crocker 1985), and there is no 
way of reproducing it assuming a Gaussian distribution of mass lost. In this case, the 
result of the Kolmogorov-Smirnov test is that the probability that a Gaussian 
distribution (with a dispersion of 0.07~$M_\odot$) reproduces observations is 0.2\%.
Hence, the case of the pair NGC~4833-NGC~5024 is not alone. 

\subsection{The correlation between mass spread and absolute magnitude}

\begin{figure}
\includegraphics[width=9cm]{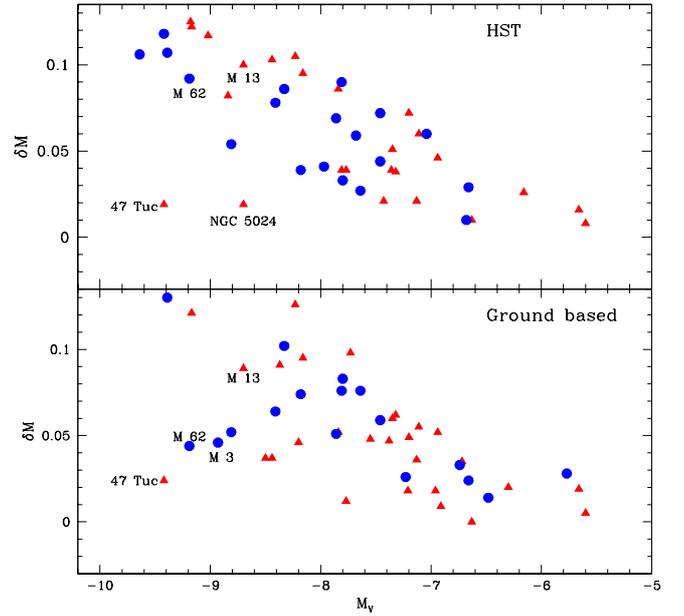}
\caption{Absolute magnitudes (M$_{\rm V}$) plotted against the difference between
the median and the minimum mass along the HB ($\delta$M). In this case, we also
separate GCs older (red triangles) and  younger (blue filled dots) than 0.92.}
\label{f:fig13}
\end{figure}

We propose that the cases discussed in the previous subsection, extracted from 
several decades of discussions, convincingly show the need for a third parameter
in addition to metallicity and age to explain the blue extension of 
the HBs. Hereinafter, we quantify the spread in mass loss by the
difference between the minimum and median mass of stars along the HB, and
call this quantity $\delta$M
\footnote{In principle, M$_{\rm min}$ should give a more correct estimate
of the spread in mass along the HB. However, we preferred
to use M$_{\rm med}$ in our discussion for the reasons explained in Sect. 2.3, 
even though in some cases it
may systematically underestimate the mass corresponding to stars
with primordial He abundances}. We note that $\delta$M has been estimated by assuming 
a constant He abundance, hence it cannot be used directly to derive the spread
in He within the cluster.

It has long been known that GCs with a large spread
of masses along the HB are bright and massive (Fusi Pecci et al. 1993). 
Recio-Blanco et al. (2006) obtained a good correlation between the 
highest effective temperature along the HB and the cluster absolute magnitude 
M$_{\rm V}$. Given that the highest effective temperature along the HB is 
well (anti-)correlated with the minimum mass of HB stars, we should expect 
a good anti-correlation between $\delta$M and M$_{\rm V}$. In 
Fig.~\ref{f:fig13}, we show the run of $\delta$M with M$_{\rm V}$ 
(M$_{\rm V}$\ values were taken directly from the Harris 1996 catalogue), 
for both the HST and ground-based data sets. There is indeed a very 
significant anti-correlation between these two quantities: large values 
of $\delta$M are only found among bright clusters. The correlation is much 
cleaner if we restrict ourselves only to those GCs with age estimates, which have the 
highest quality colour-magnitude diagrams. Hereinafter, we use these GCs only,
and combined ground-based and HST data. However, by comparing the two data
sets, we noted that the anti-correlation between $\delta$M and M$_{\rm V}$ 
is stronger when HST data are used (r=0.68 for 45 clusters) than if ground-based 
data are considered (r=0.54 for 46 clusters). This stronger correlation occurs 
because the extreme BHB stars are sometimes below the detection limit of the 
ground-based observations. Hence, when considering the two samples (whenever 
possible), we give preference to the HST data.

For this whole sample of 65 GCs (making up almost half of the total number
of Galactic GCs), the linear relation between $\delta$M and M$_{\rm V}$ is
\begin{equation}
\delta{\rm M}= -(0.102\pm 0.026) - (0.020\pm 0.003)~M_V~~~M_\odot
\end{equation}
with an extremely significant correlation coefficient of r=0.63. The r.m.s.
scatter of individual points along this mean relation (0.026~$M_\odot$) is
not much larger than expected from internal errors, suggesting that it is mainly
due to observational errors. However, part of the scatter is certainly real.
We note that there are a few clusters (namely NGC~104=47 Tuc, NGC~5024; and 
to a lesser extent NGC~362 and NGC~5272=M~3) that have a small spread of masses along 
the HB in spite of there being quite massive; one of them has indeed been used in 
the comparisons of the previous subsection. There is little doubt that 
the spread in mass along the HB is small, at least for the two most extreme 
cases. We conclude that the additional parameter required to explain the 
extension of the HB, while quite closely related to the overall cluster 
luminosity, is actually a separate one.

\subsection{He variations required explaining the HB width}

In the remainder of this section, we examine the possibility that the additional 
parameter (to both metallicity and age) determining the spread in mass for 
stars along the HB is the variation in the He abundance, related to 
various stellar generations in GCs, which combines with a small ($\sim 
0.02$~M$_\odot$) {\it random} spread in mass loss. This last value was adopted 
because it is roughly the value required to explain those GCs with the minimum 
spread in mass for stars along the HB. In principle, this value may vary 
from cluster to cluster, depending e.g., on cluster concentration (see e.g., Fusi Pecci
et al. 1993). However, we wish at present to keep our assumptions to a minimum.

To test the hypothesis presented above, we first derived the spread in 
He required to explain the observed spread in colours and masses along the HB. 
In the next section, we discuss the evidence provided by chemical 
abundances that might support this hypothesis. When deriving the spread in He, 
we should take into account that variations in He abundances have important 
effects on our analysis.

The first step of our procedure is quite simple. Since we assumed that an intrinsic
spread in the mass loss equal to 0.02~$M_\odot$ is a universal phenomenon, we
corrected the observed mass spread for this effect by subtracting this
value in quadrature from the observed $\delta$M. In those few cases where
$\delta$M$<0.02$, we simply assumed that the corrected value is 0. The
corrected spread is then attributed to a variation in the He abundance,
assuming that the same mass-loss law is applicable to all stars. 

However, to derive the variation in He, we should take into account
that the masses derived for HB stars are themselves functions of the adopted He 
abundance. Unfortunately, the Pisa evolutionary tracks were computed for a single 
(not constant) value of the He abundance for each value of the metal abundance, 
and cannot then be used to estimate this correction. We therefore 
used instead the isochrones by Bertelli et al. (2008), which while assuming a unique value 
for the mass loss along the RGB (preventing its use to derive the relation 
between masses and colours along the HB), do however provide data for different He 
abundances. We then combined the two sets of models to produce the following 
correction formula for $\Delta$M
\begin{equation}
\Delta {\rm M(Y)} = \Delta {\rm M}-\Delta {\rm Y}(1.976+1.982{\rm [Fe/H]}+0.562{\rm [Fe/H]}^2).
\end{equation}
In this equation, $\Delta$Y is the variation in He abundance with respect to
the reference value (Y=0.25, roughly the cosmological value). In this case, 
we assume that all GCs started their evolution with an He abundance close to 
that resulting from the big bang nucleosynthesis (Y$\simeq$0.25, WMAP, Spergel 
et al. 2007 --see their Table 4), but that there might be significant star-to-star 
variations even within the same GC. The procedure for estimating this He abundance 
is described in Sect.~\ref{sec:rparameter}. Of course, the primordial He 
abundance in some clusters may actually be higher than the cosmological one. 
This is supported by both simple arguments based on a rather constant 
$\Delta Y/\Delta Z$\ value throughout the Galactic evolution (Chiosi \& 
Matteucci  1982; Balser 2006; Casagrande et al. 2007), and indirect measurements 
in GC stars, e.g., those obtained using the R-method (see 
Sect.~\ref{sec:rparameter}). However, these variations in primordial He are 
expected to be quite small (a few hundredths in Y), so that this assumption is
not critical. In turn, $\Delta {\rm M(Y)}$ may be derived from the evolutionary
mass at the tip of the RGB, corrected for the uniform mass-loss law along the 
RGB. In practice, we use the equation
\begin{equation}
\Delta {\rm Y} = \Delta {\rm M(Y)}/(1.245+30.747~Z_\odot 10^{({\rm [m/H]})}),
\end{equation}
where $Z_\odot$ is the solar metal abundance, and [m/H] is the overall metallicity,
which we assumed to be [m/H]=[Fe/H]+0.28, the second additive term taking into account 
the overall enhancement of $\alpha-$elements in GCs (see Carretta et al. 2009d). 
These two equations can be solved recursively, convergence being quite fast.

Table~\ref{t:tnew7} lists the values of Y$_{\rm med}$, Y$_{\rm max}$, and  
$\delta$Y=Y$_{\rm max}$-Y$_{\rm med}$~ obtained by this procedure for each 
cluster. We recall that in some cases Y$_{\rm med}$ values might
overestimate the minimum Y in the cluster, and then underestimate $\delta$Y.
The highest He abundances found using this procedure is Y$_{\rm max}$=0.347 
for NGC~6273. The value we find for NGC~2808 is Y$_{\rm max}$=0.333; 
this value is discussed in the following section. 

\begin{figure}
\includegraphics[width=9cm]{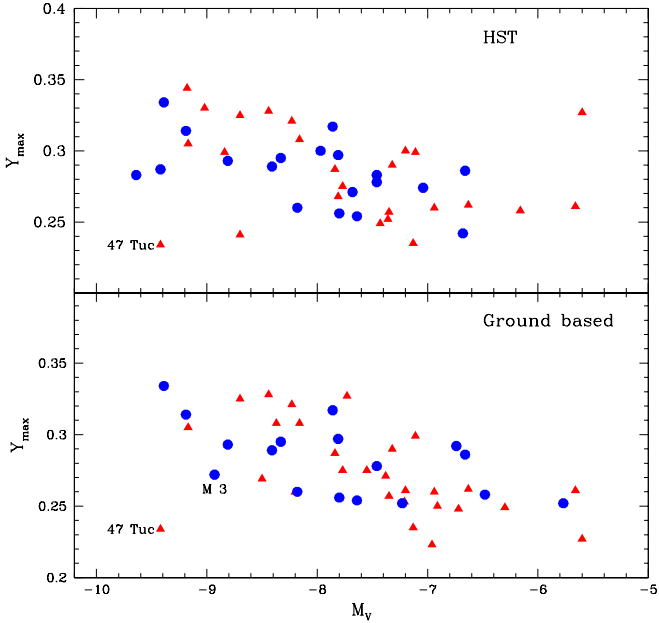}
\caption{Absolute magnitudes vs. Y$_{\rm max}$ for HST and ground-based samples;
as in all previous figures, clusters are separated according to their ages.}
\label{f:fig14}
\end{figure}

\setcounter{table}{11}
\begin{table}
\begin{small}
\caption{Spread in He and mass required explaining the HB.  The complete table is available only in electronic form.} 
\setlength{\tabcolsep}{1.5mm}           
\label{t:tnew7}      
\begin{tabular}{lcccccc}        
\hline\hline
Cluster & $\delta$M(Y) & Y$_{\rm med}$ & Y$_{\rm max}$  & $\delta$Y & $\delta~(B-V)$ & $\delta~(V-I)$\\
        &      M$_\odot$ &            &                 &           &      mag       &     mag     \\
\hline
NGC~104  & 0.000 & 0.234 & 0.234 & 0.000 & 0.000 & 0.000 \\
NGC~288  & 0.016 & 0.280 & 0.292 & 0.012 & 0.005 & 0.007 \\
NGC~362  & 0.059 & 0.243 & 0.289 & 0.046 & 0.018 & 0.025 \\
NGC~1261 & 0.068 & 0.244 & 0.297 & 0.053 & 0.021 & 0.029 \\
NGC~1851 & 0.063 & 0.247 & 0.295 & 0.048 & 0.019 & 0.027 \\
\hline                                   
\end{tabular}
\end{small}
\end{table}

In Fig.~\ref{f:fig14}, we plotted the Y$_{\rm max}$ values against the 
absolute magnitude M$_{\rm V}$. As expected, the correlation is good, yielding
\begin{equation}
{\rm Y}_{\rm max} = (0.180\pm 0.025) -(0.0134\pm 0.030) M_V
\end{equation}
with a very significant correlation coefficient of r=0.49 for 65 clusters. 

\subsection{Colour spread along the main sequence}

The spread in helium required to explain the spread in masses along the HB 
should cause a broadening of the MS. We estimated this expected colour spread 
in the MS at absolute magnitude M$_{\rm V}=+8$, by using the 
following formulae we derived by fitting data from the Bertelli et al. (2008) 
isochrones
\begin{equation}
\delta (B-V) = \delta Y (0.350-0.035 {\rm [Fe/H]})
\end{equation}
and
\begin{equation}
\delta (V-I) = \delta Y (0.672+0.099 {\rm [Fe/H]}).
\end{equation}
The colour spreads in $B-V$\ and $V-I$\ for each cluster are given in Cols.
6 and 7 of Table~\ref{t:tnew7}, respectively. Again, we recall that in 
some cases these spreads may be underestimated, because $\delta Y$ may 
itself be underestimated. This spread is very small, below detectability, for 
most of the GCs. The largest spread is expected for NGC~6273 (0.035 mag in 
$B-V$ and 0.043 mag in $V-I$); this cluster is affected by a strong differential 
reddening, which complicates the detection of this spread.

Most GCs have not yet been scrutinised in enough detail, but in a few cases
we may compare these predictions with observations. For instance, in
the case of NGC~2808, Piotto et al. (2007) found a spread of $\sim 0.1$~mag
in the $F475W-F814W$~colour from very high quality ACS data. Since we expect that
$\delta (F475W-F814W)/\delta (V-I)\sim 1.5$, the spread we predict from our 
analysis of the HB is $\delta (F475W-F814W)\sim 0.05$, which is roughly 
half the spread indeed observed. Part of this difference can be attributed 
to the median colour of the HB of NGC~2808 
($B-V_{\rm med}=0.024$) not corresponding to the red HB, but rather to 
the BHB, and in our framework is then interpreted as a moderately He 
enriched population (Y=0.273). The He-poor population is however present, 
corresponding to the red HB, which makes up almost 40\% of the cluster HB 
population. The total spread in colour along the MS is then expected to be 
larger than given simply by $\delta Y=Y_{\rm max}-Y_{\rm med}$; a more appropriate 
estimate of the expected colour spread is $\delta (V-I)\sim 0.046$\ and $\delta 
(F475W-F814W)\sim 0.07$. While this is still somewhat smaller than observed, 
the discrepancy is now small, and could be justified by some additional 
source of scatter (differential reddening, binaries, photometric errors) for 
MS colours or an incorrect calibration of the HB.

Anderson et al. (2009) found that some spread in colour is 
also present in 47 Tuc, from a very comprehensive analysis of extensive 
archive HST data. This spread is small, roughly $\delta (F616W-F814W)\sim 0.02$,
and could only be identified thanks to the exceptional quality of the data, and
the very careful procedures used to analyse them. However, the HB of 47 Tuc 
is very short\footnote{47 Tuc actually contains a few very blue HB stars 
(Moehler et al. 2000); they make up $\sim 1$\% of the HB stars in the HST 
colour-magnitude diagram we are considering. These stars are so rare that they 
do not affect our definition of M$_{\rm min}$, which excludes the lowest 5\% of 
the distribution. Indeed, this case underlines that there most likely are stars at 
the extremes of the HB whose origin is not related to their extreme values of the He
abundance.}, so that our estimate for the He spread is only $\delta Y=0.0$ (based on 
the assumption that the small residual spread in masses could be explained by 
random star-to-star variations in the mass loss adopted throughout this paper), 
and we then expect no widening of the MS due to He. A careful study of the 
HB of 47 Tuc by Di Criscienzo et al. (2009) is work in progress; early results suggest 
that it may be more accurately represented by assuming that there is a very small spread 
in Y ($\delta Y<<0.02$). By itself, this is not enough to justify the spread in 
colours of the MS. However, Di Criscienzo et al. also found that variations in 
the total CNO abundances might possibly explain both the small spread in colour 
of the MS and the far more evident split in the SGB, also found by Anderson et 
al. (2009; see however Bergbush \& Stetson 2009 for a different interpretation 
of this observation). While this comparison on the whole supports our result of 
a small spread in He in 47 Tuc, it also indicates that more careful examinations 
using synthesis of populations, taking into account the distribution in both 
colours and magnitudes of the stars on the HB, may provide additional important 
information about the properties of GCs. Extension of this careful analysis to 
many other GCs would be very helpful. 

\section{He and GC chemistry}

\subsection{Light elements anti-correlations}

Additional evidence that the spread in colours along the HB of GCs is caused by
variations in the He abundances can be obtained by considering
correlations with similar spreads in the abundances of light elements such as
Na and O, Mg and Al. As mentioned in the introduction, these spreads are
caused by different generations of stars in GCs, the ejecta
from the earliest one having polluted the material from which the
later generation(s) of stars formed. While p-captures
at high temperatures are clearly required to produce Na and Al, and destroy O
and Mg, we do not yet have a satisfactory detailed model, and even its
astrophysical basis is currently debated (either fast rotating massive stars or
massive AGB stars, experiencing hot bottom burning, or massive binaries: see 
Decressin et al. 2007; Ventura et al. 2001; De Mink et al. 2009). We are 
unable to derive the exact mass range of the polluting stars. This is an 
important concern, because the production of He has probably a different 
dependence on the mass of the polluters than both the production of 
Na and the destruction of O. For instance, if we consider the massive AGB 
scenario, He is mainly produced in the previous MS phase, and it is far 
more abundant in the ejecta of the most massive polluters (mass $>5~M_\odot$). 
On the other hand, a very significant production of Na and depletion of O can 
be obtained even within stars of lower mass. Hence, the ratio of He
to Na production might change from cluster-to-cluster, provided that the
mass range of polluters changes. With this caveat in mind, we can then
examine current evidence.

In principle, the original He content should be attainable from direct 
measurements for each HB star. However, this datum is neither available nor can 
be easily obtained, save possibly for a restricted temperature range at 
$\sim 10000$~K (Villanova et al. 2009; He is heavily depleted by 
sedimentation in warmer HB stars: see e.g., Behr et al. 2000b, Behr 2003). 
Furthermore, even O and Na abundances are not available for most HB stars 
(for a possible exception, see again Villanova et al. 2009), although in this 
last case abundances could in principle be obtained for stars on the RHB, 
within the instability strip, and on the coolest part of the BHB. Because of
this shortage of data, we use values provided by RGB stars. Of 
course, the implication is that only statistical properties of the distributions, 
not individual values can be studied. In practice, we only consider extreme 
values, with the assumption, consistent with our approach, that the bluest 
(i.e., less massive) HB stars are the progeny of the most He-rich TO-stars, and 
the reddest ones of the most He-poor ones; in a future paper, we plan to consider 
in more detail other characteristics of these distributions.

\begin{figure}
\includegraphics[width=9cm]{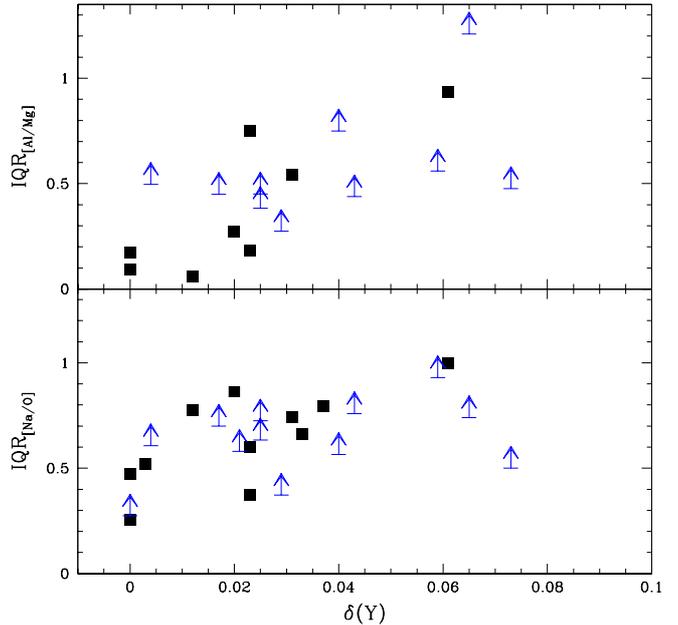}
\caption{$\delta$Y as function of IQR for [Na/O] and [Al/Mg], respectively. Lower
limits are also indicated.}
\label{f:dyiqr}
\end{figure}

In the present discussion, we considered a subset of the GCs with
extensive enough data on the Na-O and Mg-Al anticorrelations. Most of the
GCs considered here are from the very extensive study by Carretta et al.
(2009a, 2009b); we complement this data set with  a few GCs from the
literature (Shetrone \& Keane 2000 for NGC~362; Sneden et al. 2004 and
Cohen \& Melendez 2005 for NGC~5272 and NGC~6205; Sneden et al. 1991, 2000
for NGC~6341; Marino et al. 2008 for NGC~6121; Marino et al. 2009 for
NGC~6656; and Yong et al. 2005 for NGC~6752). The relevant data are given in
Table~\ref{t:tnew0}. We note that we prefer to use the inter-quartile IQR 
of the distributions (either directly taken from the literature, or obtained 
from abundances for individual stars\footnote{We thank A.F. Marino for
providing the unpublished data for the individual stars in NGC~6656.}), 
rather than the corresponding values for 90\%
of the distribution, as we did for the extension of the HB. The IQR
is indeed a more robust indicator, which is less sensitive to small number
statistics and to many upper limits to the abundance
determinations. This last problem remains an important concern.
In general, we may consider the IQR determinations as lower limits for
those GCs with metallicity [Fe/H]$<-1.5$. Figure~\ref{f:dyiqr} illustrates the
correlations between the spread in He abundances obtained from the colour
spread of the HB, and the IQR values for both the [Na/O] and the [Al/Mg]
anticorrelations. As can be seen, fairly tight correlations exist,
the strongest being between $\delta$Y and IQR([Na/O]), which are based
on more extensive data sets than the IQR([Al/Mg]). This correlation
strongly supports the current interpretation that the extent of the HB is
determined by the spread in He content within each GC. However, we note
that at least for $\delta$Y-IQR([Na/O], the relation has an
offset, where $\delta$Y is significantly larger than 0 only for clusters with
IQR([Na/O]$>0.6$).

\subsection{Comparison with a dilution model}

\begin{figure}
\centering
\includegraphics[bb=30 170 570 510, clip, width=9cm]{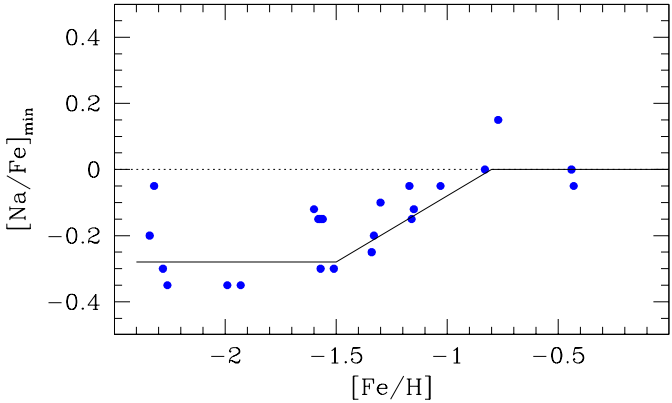}
\caption{Run of [Na/Fe]$_{\rm min}$\ with [Fe/H] for the GCs of our sample.
The thick superimposed line represents an average line fitting field stars (see text) }
\label{f:nafe}
\end{figure}

We may compare these results with the prediction of a simple universal
mechanism for the production of the Na-O correlation, that has only one class of 
polluters, and for which the composition of second generation stars differ only in terms of
the different dilution of polluted gas by primordial material. A similar 
dilution model has been successfully used to explain many features of the 
Na-O anti-correlation (see discussion in Prantzos \& Charbonnel 2006), 
including its shape and the residual observed Li abundances in stars 
severely depleted in O in NGC~6752 (Pasquini et al. 2005). Once the 
compositions of the pristine and processed material are set (e.g., by the
extremes of the observed distributions), the dilution factor may be
determined for each star (either from O or Na abundances), and the
helium production can be inferred, save for a constant factor.

In this model, the logarithmic abundance of an element [X] for a given
dilution factor $dil$ is given by
\begin{equation}
[X] = \log{[(1-dil)~10^{\rm [X_o]} + dil~10^{\rm [X_p]}]},
\end{equation}
where [X$_o$] and [X$_p$] are the logarithmic abundances of the element 
in the original and processed material, respectively. Practically speaking, 
we may assume (see Carretta et al. 2009a, Fig. 17) that the original
composition is [Na/Fe]=-0.28 if [Fe/H]$<-1.5$, [Na/Fe]=0 if [Fe/H]$>-0.8$,
and [Na/Fe]=-0.28+0.4([Fe/H]+1.5) if -1.5$<$[Fe/H]$<$-0.8; and [O/Fe]=0.5.
For the processed material we assumed [Na/Fe]=0.6 and [O/Fe]=-1.5. We note
that while for O we may assume a uniform [O/Fe] value, because
cluster-to-cluster variations of maximum values are small among the GCs
of our sample, we adopted an original [Na/Fe] ratio that is a function of
metal abundance, to reproduce both our data and the Na abundances
observed among field stars (see Fig.~\ref{f:nafe}).

\begin{figure}
\centering
\includegraphics[bb=30 170 570 510, clip, width=8.5cm]{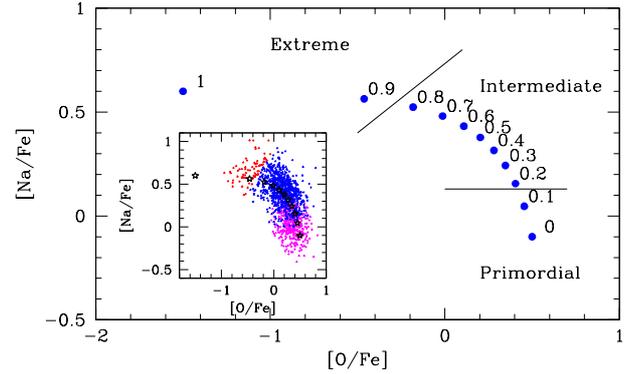}
\caption{Run of [Na/Fe] vs. [O/Fe] expected for different values of $dil$\ from our
dilution model. The insert shows the run of [Na/Fe] with [O/Fe] for stars in
our FLAMES survey (Carretta et al. 2009a).}
\label{f:dilution}
\end{figure}

In Fig.~\ref{f:dilution}, we compare the expected run of [O/Fe] vs.
[Na/Fe] according to this dilution model, with observations for the
ensemble of stars observed within our project (Carretta et al. 2009a). It
is important to notice that within this approach the production of Na (and
of He) saturates at large values of the dilution parameter {\it dil}. This
has two important implications: (i) most reliable estimate of the maximum
value of {\it dil} is given by the O abundances; and (ii) that similar He
abundances are obtained for a rather wide range of [O/Fe]. On the other
hand, the minimum value of {\it dil} is most accurately given by Na abundances,
because similar values of [O/Fe] are obtained for small values of {\it
dil}. Minimum and maximum values of O and Na abundances measured in each
cluster are given in Cols. 2, 3 and 4 and 5 of Table~\ref{t:t1}. We note
that while Na and O maximum abundances listed in this Table are somewhat
different from the values adopted in our dilution model, this has no
consequences on our discussion because minimum and maximum values of the
$dil$\ parameter (Cols. 6 and 7 of Table~\ref{t:t1}) are obtained from
the minimum abundances of Na and O, respectively. Both minimum and maximum
values of {\it dil} contain some uncertainty, because only upper limits to
O and Na abundances can be
obtained in extreme cases; this caveat is stronger for metal-poor stars,
and for the smallest GCs, where most of the observed stars are quite warm.
For instance, we suspect that the minimum Na abundance is overestimated in
the case of M~15, which is a very metal-poor GC, and that for this reason
we underestimate the $\delta$~{\it dil} range appropriate for this cluster.
Finally, the differences between the maximum and the minimum dilution are given 
in the last column of Table~\ref{t:t1}. Within this model, these values should 
be proportional to $\delta$Y. 

\setcounter{table}{12}
\begin{table}
\setlength{\tabcolsep}{1.3mm}
\caption{Minimum and maximum values for [Na/Fe] and [O/Fe] (see text for references). 
The dilution fraction is also reported.}		
\label{t:t1}	 
\centering			   
\begin{tabular}{lccccccr}        
\hline\hline
NGC~& \multicolumn{2}{c}{[Na/Fe]} &\multicolumn{2}{c}{[O/Fe]}&
\multicolumn{2}{c}{$dil$} & $\delta ~dil$ \\
&min&max&min&max&min&max & \\
   &     dex    &   dex & dex & dex
  &   & &      \\
\hline
104  & 0.15& 0.74&-0.40& 0.38&~0.14 & 0.87 &0.73  \\
288  &-0.10& 0.71&-0.50& 0.36&~0.09 & 0.90 &0.81  \\
362  &-0.15& 0.40&~0.10& 0.38&-0.00 & 0.51 &0.51  \\
1904 &-0.15& 0.72&-0.60& 0.28&~0.12 & 0.93 &0.81  \\
2808 &-0.12& 0.56&-1.00& 0.37&~0.02 & 0.99 &0.97  \\
3201 &-0.30& 0.60&-0.80& 0.32&-0.02 & 0.97 &0.99  \\
4590 &-0.35& 0.53& 0.00& 0.72&-0.05 & 0.62 &0.67  \\
5272 &-0.15& 0.55&-0.10& 0.50&~0.12 & 0.70 &0.58  \\
5904 &-0.25& 0.60&-0.70& 0.43&-0.03 & 0.95 &0.98  \\
6121 &-0.05& 0.74&-0.20& 0.37&~0.08 & 0.77 &0.69  \\
6171 &-0.05& 0.69&-0.30& 0.39&~0.03 & 0.83 &0.80  \\
6205 &-0.12& 0.70&-1.00& 0.50&~0.15 & 0.99 &0.84  \\
6218 &-0.20& 0.67&-0.40& 0.56&~0.01 & 0.87 &0.86  \\
6254 &-0.30& 0.56&-0.40& 0.47&-0.02 & 0.87 &0.89  \\
6341 &-0.30& 0.45&-0.10& 0.38&-0.02 & 0.70 &0.72  \\
6388 & 0.00& 0.67&-0.60& 0.24&~0.00 & 0.93 &0.93  \\
6397 &-0.35& 0.71&~0.00& 0.37&-0.05 & 0.62 &0.67  \\
6441 &-0.05& 0.80&-0.40& 0.20&-0.04 & 0.87 &0.91  \\
6752 &-0.15& 0.65&-0.40& 0.53&~0.12 & 0.87 &0.75  \\
6809 &-0.35& 0.69&-0.20& 0.44&-0.05 & 0.77 &0.82  \\
6838 & 0.00& 0.76&~0.00& 0.48&~0.01 & 0.62 &0.61  \\
7078 &-0.05& 0.70&-0.10& 0.49&~0.23 & 0.70 &0.47  \\
7099 &-0.20& 0.76&-0.20& 0.60&~0.07 & 0.77 &0.70  \\
\hline					
\end{tabular}
\end{table}

We note that while this simple dilution model predicts quite uniform 
values of $\delta~dil$, and then $\delta Y$, the values we derived from the 
HB exhibit large variations from cluster-to-cluster. As a consequence, the
correlation for individual clusters is poor. We may also remind that this
simple model, which uses a universal polluter, also predicts that the
Na-O and Mg-Al anticorrelations should closely resemble each other, which is not
observed (see Carretta et al. 2009b).

We conclude that while a second generation of stars polluted by some stars
of an earlier generation exists in all GCs, the composition of the
polluters is not universal. In some clusters (such as NGC~2808), the polluters 
produce large amounts of He (most likely with similarly large amounts of 
Al), and very efficiently destroy O and Mg; while in others (like NGC~6121) 
practically no fresh He (and Al) is present in the material that produced the 
second generation stars. On the other hand, all these polluters produce 
similar amounts of Na. Independent of the polluter, we should be able to reproduce 
these observations, within a viable scenario for cluster formation and early 
evolution.

\subsection{The impact of cluster luminosity}

\begin{figure}
\includegraphics[width=9cm]{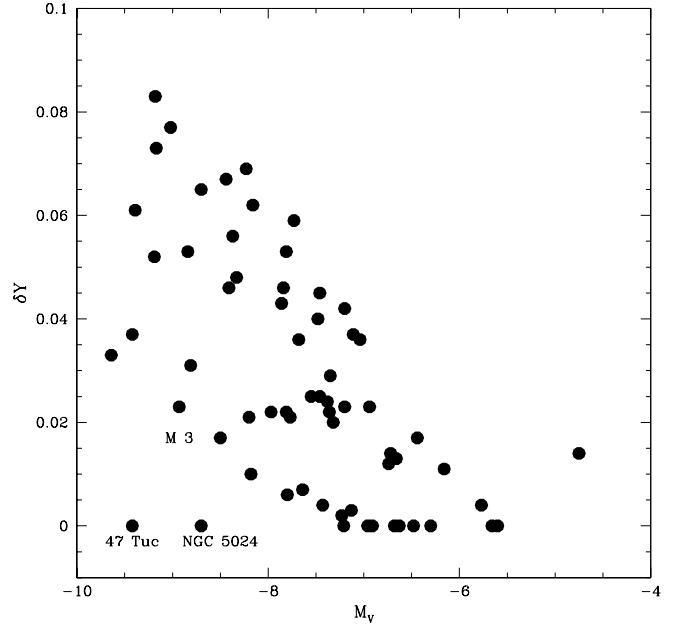}
\caption{Run of $\delta$Y with absolute magnitude M$_{\rm V}$.}
\label{f:dYmv}
\end{figure}

The previous discussion suggests that the spread in the He content within stars
in GCs are the most important factor determining the extension of the
HBs. It is important however to understand the basic physical
property of GCs causing this spread in He. The existence of a good
correlation between the maximum temperature of the HB and the cluster
luminosity (Recio-Blanco et al. 2006) suggests that the total cluster
mass determines its ability to have stars of very different He
contents, assuming that the present cluster luminosity is a good proxy for its 
original mass. In Fig.~\ref{f:dYmv} we show the correlation that we find between 
the spread in He abundance (as represented by $\delta$Y) and cluster 
luminosity $M_{\rm V}$, here a proxy for the total mass. There is a clear 
strong trend of an increase in $\delta$Y with decreasing $M_{\rm V}$ (that 
is an increase in luminosity). The linear correlation coefficient is 0.56 
for a total of 65 clusters, which is significant at a level of more than 99.99~\%. 
The mean regression line is
\begin{equation}
\delta Y = -(0.0127\pm 0.0024) M_V - (0.069\pm 0.020).
\end{equation}
Although the correlation between the spread in He abundances (derived from
the width of the HB) and luminosity is very good, there are a few
clear exceptions, which are massive clusters with very short HBs (and hence
small values of $\delta$Y), the clearest examples being NGC~104 (47 Tuc),
and NGC~5024, and to a lesser extent NGC~5272 (M~3). We emphasize that the deviation 
of these clusters from the trend shown by the vast majority of the clusters 
is much larger than observational errors. It is notable that two of the most 
well studied clusters (47 Tuc and M~3), often considered as templates of metal-rich 
and metal-intermediate GCs for several studies of the HB, are found to be 
the exceptions rather than the rule. We recall that both
NGC~104 and NGC~5272 have rather small values of IQR[Na/O], much less than
observed in other clusters of similar luminosity. This again emphasizes the
connection existing between the extension of the HB and the Na-O
anticorrelation, hence supporting the explanation we suggest that the width of
the HB is related to spreads in the He abundances.

We finally recall that D'Ercole et al. (2008) and Carretta et al. (2009d)
presented a scenario for the formation of GCs in which the
correlation between cluster luminosity and extension of its HB may occur
naturally. According to that scenario, the formation of a GC is the final
act of a series of events that begins from the violent onset of star
formation in a very large original cloud. Star formation in this cloud
continues until the kinetic energy injected by SNe and massive star
winds causes dissipation of the remaining gas. In this scenario, this
early population is not compact enough for the formation of a stable
cluster, and should dissipate after the violent relaxation produced by the loss
of the remaining gas and the ejecta of the most massive stars. However,
once this very violent initial phase terminates, a much quieter
situation follows, where there is no longer a strong injection of
energy in the ISM; the slow winds from stars of intermediate mass may then lead
to the onset of a cooling flow, and a formation of a very dense and
kinematically cold cloud, from which the present GC may form, mainly
composed by second generation stars (the I and E populations of Carretta et
al. 2009a). Part of the primordial stellar population remains trapped in
the current GCs, and forms what Carretta et al. (2009a) called the
P-population. In this scenario, it is quite natural to expect that the
onset of star formation within the cooling flow should occur earlier in
more massive clusters; it will also probably stop earlier, because of the
kinetic energy injected by the massive stars of this second generation.
Therefore, we may expect that the delay of the second generation, hence
the typical polluter mass, is roughly determined by the mass of the GC
itself. A few exceptions may be easily accommodated in this scenario as
objects characterised by a prolonged phase of formation of the primordial
population, e.g., caused by the presence of a very extended region of star 
formation or companion clusters, as often observed among LMC populous 
clusters.

This scenario is clearly constructed on the hypothesis of massive AGB stars 
being polluters. Were rotating massive stars the polluters, some important 
modifications would be required, because these stars lose most of their 
mass in the epoch of core-collapse SNe explosions, and it would be very 
difficult to produce the required cooling flow. Reproducing the trends 
considered in this paper would clearly be a challenge for this alternative 
scenario.

\section{The M~3-M~13 pair}

Since the paper by Van den Bergh (1967), the pair M~3-M~13 (=NGC~5272 and
NGC~6205) has played a very special r\^ole in the second parameter issue. 
This pair has been repeatedly studied to search for evidence 
of He-abundance variations (see Caloi \& D'Antona 2005, and Catelan et al. 2009). 
They are bright, massive and amongst the easiest GCs observable from the 
northern hemisphere. M~3, with its large population of RR Lyrae, has 
often been considered the archetype of Oo~{\sc i} GCs, while M~13 is considered 
to play the same r\^ole amongst the clusters with very blue HB. It is interesting 
to revisit them within the scenario outlined in this paper.

According to the basic parameters defined in this series of papers, M~3 and
M~13 might not be so close twins, after all. While the metal abundance is 
quite similar ([Fe/H]=-1.50 and -1.58), the ages might be different. Without
considering the correction for the different He abundance, the relative ages 
we considered are 0.88 and 1.02 (see Table~1). After this correction is
taken into account, this difference becomes even a bit larger (0.875 and 
1.033). It is then possible that
M~3 is younger then M~13, a result found in some careful determinations 
(see Rosenberg et al. 1999, and De Angeli et al. 2005), but not in others 
(Mar\'in-Franch et al. 2009; Dotter et al. 2010). This by itself might explain 
why the median colour of the HB of M~13 is much bluer than that of M~3, without any 
need for a very large difference in Y$_{\rm med}$ (we actually obtained values of 
0.249 for M~3 and 0.260 for M~13).

However, it has for a long time been known that these two GCs also have a large 
difference in the extension of their HBs. According to our analysis, the spread 
in mass along the HB of M~3 is small. Assuming a constant helium abundance,
we obtain minimum, median and maximum masses of 0.624$\pm$0.003, 0.670$\pm$0.001, 
and 0.716~M$_\odot$, respectively. Apart from a general offset of about 
0.02~M$_\odot$ (which resembles the overall offset of masses derived from HST and 
ground-based data in our analysis), these values are quite similar to those obtained
in the analysis of Valcarce \& Catelan (2008), although it is possible that some
very blue/faint stars considered by these authors were actually
not studied by the Rosenberg et al. (2000a) photometry. As discussed by several
authors, the distribution of masses along the HB is probably not Gaussian (see 
Castellani et al. 2005 and Valcarce \& Catelan 2008). If the interpretation that we
give for the mass spread along the HB in terms of variations in the He
content is correct, we would correct the offset between median and minimum
mass to 0.030~M$_\odot$, and the corresponding spread in helium would be
$\delta$Y=0.023. Catelan et al. (2009) argued that such a small spread in He
in M~3 would be detectable from Str\"omgren photometry and spectroscopy of the
blue HB stars, and concluded that a uniform He abundance more closely matches the observations
for this cluster. This might suggest that small spreads in He such as the one we
derived for M~3 might be an artifact of a non-Gaussian {\it random} mass-loss
distribution. However, since the bluest HB stars ($T_{\rm eff}>11,500$~K, consisting of 
some 5\% of the whole HB population of M~3) could not be analyzed by Catelan et al., 
it is not entirely clear that there is no real spread in helium within M~3.

On the other hand, M~13 has a very wide range of both colours and masses along the HB. 
The spread in mass and Y abundances we obtained using $M_{\rm med}$ are 
$\delta$M=0.082~M$_\odot$ and $\delta$Y=0.065. Had we instead used $M_{\rm max}$,
scaled down by 0.02~M$_\odot$\ to take into account the random term in mass lost,
we would have got $\delta$M=0.107~M$_\odot$ and $\delta$Y=0.083. We expect then
a detectable spread in colours for the MS of M~13 [$\delta (V-I)=0.033$~mag and 
possibly as much as $\delta (V-I)=0.043$~mag, if $M_{\rm max}$ were adopted,
assuming that the minimum Y value is smaller than Y$_{\rm med}$. This colour
spread could possibly be detected in the future by a careful analysis of HST 
photometry.

As noticed by several authors, this huge difference in the width of the HB
between M~3 and M~13 is quite unexpected, since M~3 is a luminous cluster 
($M_V=-8.93$), even more luminous than M~13 ($M_V=-8.70$). As the discussion 
in Sect. 6.4 shows, we should indeed expect a more extended HB for a 
luminous cluster like M~3. As a matter of fact, M~3 (and moreover 47 Tuc and NGC~5024) 
appears peculiar with respect to the other GCs, while M~13 appears to be similar to the 
majority of GCs in this respect. Within the scenario we described in the 
previous section (see also Carretta et al. 2009d), we are tempted to invoke a 
delayed cooling flow to explain the peculiarity of M~3 and 47 Tuc. This might 
explain other differences between M~3 and 47 Tuc on the one hand, and the other 
massive GCs on the other, such as the small values of IQR[Na/O]. However, we 
admit that without any further information, this hypothesis might look rather ad hoc
\footnote{After the first draft of this paper was written, 
Lane et al. (2010) suggested that 47 Tuc might have formed by the merging of a 
binary cluster. This suggestion comes from a completely different line of thought, 
being based on the internal kinematics. As mentioned in the previous section, 
it is possible to speculate on a possible connection between such an origin
of 47 Tuc and its anomalously short HB.}. Thus, ironically, while the present
approach seems to explain most of the observations relate to the second
parameter effect, its application to the most classical couple M~3-M~13 still
encounters difficulties.

\section{Conclusions and suggestions for further work}

As we have seen, our reanalysis of public extensive photometric databases 
of GCs demonstrates that {\it age is the main second parameter} affecting 
the HB morphology. This hypothesis is able to explain quite well most of the observables related to 
median HB stars of GCs, when coupled with a simple mass-loss law that is a 
linear function of [Fe/H]. Among the many observables that are successfully 
explained, we note the Oosterhoff dichotomy that we attribute to the 
peculiar age-metallicity distribution of Galactic GCs. Oo~{\sc ii} clusters 
are mostly old, while Oo~{\sc i} are predominantly young, although 
young Oo~{\sc ii} and old Oo~{\sc i} GCs exist. However, {\it at least a third parameter 
is required} (and possibly even others) to fully explain the median colours 
of HBs (in particular those with very blue HBs) as well as their extension. 
There are various reasons to identify this third parameter with {\it variations 
in the He content}. These include the variation in the scatter with metallicity, some 
correlation with the R-parameter, and the clear links with chemical anomalies 
observed in GCs. This result is strongly indicative of a possible link between the 
colours of the stars on the HB and their original composition, in a multiple 
generation scenario for the formation and early evolution of GCs. Self-pollution 
in GCs is possibly responsible for a large variety of the second parameter features,
and may be in part described using the Na-O anticorrelation, although some 
modulation according to cluster luminosity is required.\footnote{The scenario we
propose should of course not only explain the Milky Way GCs, but e.g., those of Fornax
(Buonanno et al. 1998). This case is quite puzzling, since clusters 1, 3, and
5 are nearly coeval, and have similar metallicity, and still have very different 
HB's. We did not quantify these variations in terms of mean colours and magnitudes 
as done for the Milky Way GCs considered in the present paper, hence we cannot 
provide any quantitative analysis. We only note that the ranking of HBR ratios for the 
three coeval clusters of similar metallicity (1, 3, and 5: -0.2, 0.50, and 0.44)
is the same as the ranking in absolute magnitude $M_V$ (-5.32, -7.66, -6.82).
A quantitative analysis is required to settle this issue.}

A combination of age and He variation therefore appears to be an explanation of the 
long-standing problem of the second parameter, although we do not exclude
additional parameters such as the CNO abundances or the presence of binaries (e.g.,
of the progeny of blue stragglers) possibly playing some r\^ole. However, this issue is 
still far from being completely settled. We need to make some progress in developing
models, and a number of observational tests. A short list includes:
\begin{itemize}
\item Understanding the nature of the polluters. This requires advances in the
modeling of AGB stars and rotating massive stars. Furthermore, detailed 
spectroscopic data for stars in massive and very young clusters, such as RSGC1 and 
RSGC2 (Davies et al. 2007), or intermediate age clusters in the LMC, 
where multiple MS TO's have been observed (Mackey et al. 2008; Milone et al. 2009), 
may provide a crucial test of this scenario.
\item The present discussion has focused on He, Na and O, but the abundances of other elements 
may also play an important r\^ole. For instance, Al might be a proxy for He that is more reliable
than Na. Unfortunately, our data for Al are not as extensive as those for Na and O, but
the relation between the production of He (which most likely occurred within MS stars),
and the proton capture processes (which might have occurred in the same main sequence
stars, if massive and fast rotating, or later during the AGB phase, if the stars were
of intermediate mass) must be clarified.
\item A number of confirmations of this scenario are required. These include (i) direct 
determinations of He, Na, and O in HB stars, which were shown to be possible in at least
some cases by Villanova et al. (2009); and (ii) a discussion of the luminosity of the RGB 
bump that we defer to another paper currently in preparation.
\item In addition, we ask: do properties of RR Lyrae variables agree with expectations? Are 
anti-correlations found where expected (important clusters such as M~54 and NGC~1851 
do not yet have adequate data)? Are multiple sequences observed where they are expected? 
We note here that while the connections between multiple MSs and variations in the He 
abundance is quite clear, the case of multiple sub-giant branches (SGB) is more ambiguous. 
SGB splitting measured using visual-red-near infrared colours (see e.g., Milone et al.
2008) might be due to a variation either in age or most likely total CNO content
(see e.g., D'Antona et al. 2009 and Cassisi et al. 2008), or even [Fe/H] (in this case 
however some spread in the MS and RGB is also expected). Interpretation of splitting is even
more ambiguous when considering ultraviolet colours, which are sensitive to N excesses. 
Variations in He abundances are only marginally effective in these cases, because sequences differing 
only in Y are very close each other on the SGB (D'Antona et al. 2002). Variations in total 
CNO content can be most likely attributed to the contribution of thermally pulsing AGB stars 
(Cassisi et al. 2008), 
which have a rather low mass and probably do not contribute much to He abundance variations. 
It is then unclear that there should be any correlation between the SGB splitting and 
large spreads on the HBs. In fact, NGC~2808 has a quite narrow SGB (Piotto et al. 2007). 
SGB splitting has been detected using visual-red-near infrared colours in 47 Tuc 
(Anderson 2009), NGC~1851 (Milone et al. 2008), and NGC~6388 (Moretti et al. 2009). These 
clusters have very different HB morphologies, ranging from very short (47 Tuc), to 
bimodal (NGC~1851), to very extended (NGC~6388). We obtain very different estimates 
of the He spread (0 for 47 Tuc; 0.048 for NGC~1851, a possibly too large value compared 
with those determined by Salaris et al. 2008, and Catelan et al. 2009b; 0.037 for NGC~6388). 
This lack of correlation suggests that the two phenomena are somewhat different, as 
expected if the mass range of the polluters changes from cluster-to-cluster.
\item The scenario requires a number of refinements. All analytic dependencies 
adopted throughout this discussion should be reviewed, and possibly replaced by a
comparison with synthetic HBs. This may allow us to detect additional effects,
e.g., a variation in total CNO abundances, not included in the present analysis.
\item Finally, hydrodynamical simulations of the formation and early evolution of
massive star clusters are urgently needed. There are aspects of the
present scenario that are necessary to explain observations, but should be understood
more clearly. The most intriguing is the existence of a pool of gas from which second
generation stars formed, which is composed of material processed through H-burning at 
high temperature diluted with pristine gas. How this pool of gas is generated, and
how the stars form from it within the potential well of the young GC, while
other stars of the original population evolve, remains unclear. Some explorative 
results were obtained by D'Ercole et al. (2008) which while very promising
should be placed on a sounder basis.
\end{itemize}

\begin{acknowledgements}
The authors wish to thank the anonymous referee for insightful 
and useful comments which lead to a significant improvement of the paper.
L. Girardi for having provided the results of the TRILEGAL simulations used
for field decontamination.
F. D'Antona and S. Cassisi deserve to be acknowledged for a critical reading of the manuscript
and for the very valuable suggestions. 
Finally, we thank Mariangela Bonavita for help in the preparation of the
figures, and the Italian MUR for financial support through PRIN 20075TP5K9.
This research has made use of NASA's Astrophysics Data System.
\end{acknowledgements}


\end{document}